\documentclass[intlimits,twoside,a4paper]{article}
\usepackage{amsmath,amssymb}
\usepackage{graphicx}
\usepackage{grffile}
\usepackage{wrapfig}
\usepackage[T2A]{fontenc}
\usepackage[cp1251]{inputenc}

\usepackage[eqsecnum]{cmpj2}
\issue{2012}{15}{2}{23001}
\doinumber{10.5488/CMP.15.23001}

\title[Ion clusters and ion-water potentials]%
      {Ion clusters and ion-water potentials \\ in MD-simulations\thanks{Dedicated
      to Dr. Orest Pizio on the occasion of his 60th birthday.}}
\author[Ph.A. Bopp, K. Ibuki]
{Ph.A. Bopp\refaddr{label1}, K. Ibuki\refaddr{label2}\thanks{Professor Kazuyasu Ibuki passed away
on May 22, 2012.}}
\addresses{
\addr{label1} {Universit\'e de Bordeaux, Department of Chemistry, \\
351 Cours de la Lib\'eration, B\^at. A12, FR--33405 Talence cedex, France}
\addr{label2} {Doshisha University, Department of Molecular Chemistry and Biochemistry,
 Kyotanabe, Kyoto 610--0321, Japan
}}

\authorcopyright{Ph. A. Bopp, K. Ibuki, 2012}

\date{Received  December 31, 2011, in final form  March 13, 2012}

\begin{document}

\maketitle

\begin{abstract}

A well known, if little documented, problem in many molecular simulations of
aqueous ionic solutions at finite concentrations is that unrealistic
cation-cation associations are frequently found. One might suspect a defect in the
ion-ion interaction potentials, about which not much is known. However, we
show  that this phenomenon can also be traced to the fact that, in the
pair-potential approximation, the cation-water potentials are too deep
compared with the other ones and we investigate this phenomenon in some detail. We
then attempt to draw some general conclusions.
\keywords molecular dynamics simulations, aqueous ionic solutions,
ion association

\pacs 02.70.-c, 02.70.Ns, 61.20.Ne, 61.20.Qg, 61.20.Gy, 61.20.Ja

\end{abstract}

\section{Introduction}

It is a truism to stress the importance of aqueous salt solutions for
many fields of science. Enormous efforts have consequently been made
in the past decades to go beyond phenomenological considerations and
to gain a fuller understanding of their structure
and dynamics at the molecular level. A great many strategies
in theoretical and computational physics and chemistry have been deployed
for this purpose.

It turns out that while the hydration of almost all  simple positive ions has
by now been investigated in Molecular Dynamics (MD) simulations at many levels
of approximation (many pair  potentials (see e.g.~\cite{liem1}), three body
potentials~\cite{Lauenstein2000},  various QM-MM methods (see e.g.~\cite{rode2}
and references therein), Car-Parinello
MD (see e.g.~\cite{Pollet2007}), Brownian dynamics and related approaches~\cite{Jardat1999}),
studies of solutions at finite concentrations are much scarcer. This is of
course due to the much more difficult modeling task that one faces in this
case.  At least six interaction potentials (ten pair potentials in this
approximation) must be determined in a consistent way. Their relative
contributions to the total energy will vary widely when the salt concentration
is varied. Any small imbalance between various terms can then lead to
artefacts. We will come back to this point below.

\begin{figure}[ht]
\centerline{
\includegraphics[width=5.8cm]{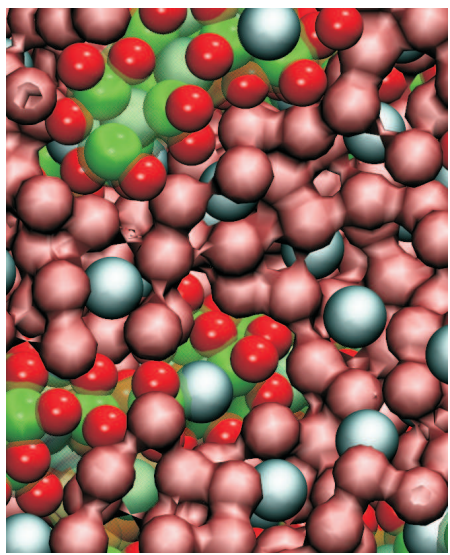}   % width=7.5cm
}
\caption{(Color online) Snapshot of a typical local arrangement in a
10~m LiCl solution at 300~K. Flexible water, $\alpha=1$, see text.
The Li$^+$-ions are dark green,
the Cl$^-$-ions blue,
water molecules closer than  2.5~{\AA} to a Li$^+$-ion (hydration shell) are
bright red, other water molecules pink.  Hydrogen atoms are
omitted. The Li$^+$-clusters together with the hydration water and
Cl$^-$-ions close to Li$^+$-ions (ion pairs) have been joined
together by a green
surface, the `bulk' water by a brownish one.}
\label{fig1}
\end{figure}

We have recently studied  aqueous LiCl solutions in their entire concentration
range at 300~K and normal densities~\cite{Ibuki2009} using a well established
class of models that had been used  for many salt solutions at low
concentrations. Among the findings we observed at intermediate concentrations 
long-lived (with respect to the simulation times of a few 100 picoseconds)
aggregations of Li$^+$-ions. Figure~\ref{fig1} shows such a situation as it is
found in a 10~m solution. Partly hydrated Li$^+$-ions aggregate, leaving thus
space for small, possibly interconnected, `pools' of water with the weakly
solvated Cl$^-$-ions.  The shape of these cationic aggregates is generally more
or less linear and the ions are either in direct contact or barely solvent
separated, as seen in the figure.

%fig. 1

The structural details of very concentrated ionic solutions
at room temperature are
little known. There is some evidence for the existence of, albeit
possibly small (down to 5 H$_2$O molecules),
nano-pools of water or worm-like
structures in LiCl solutions at low temperatures~\cite{Turton2011}.
Aggregations of some cations have also been found to
influence the formation of certain complexes
in aqueous solutions
at room temperature~\cite{Terekhova2010}.
Even though they can thus not be entirely excluded,
we do not know of any
convincing experimental evidence for the existence of
such aggregates, concerning either the ions or the
solvent, in bulk solutions
at higher temperatures. We therefore, decided to
consider the Li$^+$ aggregates mentioned above
as a possible artefact of our model
and to study them in some
detail in an attempt to trace their existence to some feature of
our interaction model. We report here the results of this analysis
and new simulations of the same solutions
where this phenomenon has been eliminated by
various means.

\section{Interaction models, general considerations}

The interaction models used in simulations of salt solutions are usually built
up by combining a model which has been developed and tested for pure water
(e.g. ST2, MCY, SPC, SPCE, TIP4P, BJH) with ion-water potentials developed in
various ways, mostly by fitting under different constraints (e.g., the charge of
the ion and geometry and partial charges of the water model) empirical
functions to ab-initio energies. Simple charged Lennard-Jones (LJ) spheres
are mostly, also in this work, assumed for the ion-ion interactions.

Besides the problem discussed in the introductory section, several other ones
have been identified. There are many critical discussions of the models for
pure
water~\cite{Groenenboom2000,Mas2000,Guillot2001,Guill02,Pusztai2008}, but
we shall not enter this vast debate here. Pusztai et al.~\cite{Harsanyi2011}
have in particular concluded from comparisons of scattering experiments,
reverse Monte Carlo (rmc), and MD simulations that, at higher salt
concentrations, the water-water interactions developed for pure water may no
longer be adequate.

There is indeed no guarantee whatsoever that interaction potentials
of so different origins, often
involving different approximations (effective potentials), can be combined
in (almost) arbitrary proportions, depending on the salt
concentration. Studying single ions, or solutions
at low concentration, alleviates this problem. However,
most `real' solutions (biological fluids, sea water, in industry)
are  such that concentration effects cannot be neglected.

In particular, an imbalance between the cation-water and water-water interaction
potentials could lead to artefacts above a certain concentration, i.e. when the
ion-water energies are no longer small compared to e.g. the water-water ones.
It can then happen that it becomes energetically more favorable to accept some
ion-ion repulsion (especially between monovalent ions) and somehow allow these
ions to associate so that they can  share the remaining (scarce) water
molecules to form a common `hydration shell' for some sort of  M$_n^{n+}$
aggregate.  We explore this phenomenon (see e.g. in~\cite{Ibuki2009}) here by
modifying the Li$^+$-water interaction employed in~\cite{Ibuki2009} in two
different ways: First by reducing the Coulomb-terms in the Li$^+$--O  and
Li$^+$--H pair potential, and secondly by rigidifying in different ways the
water molecule.

\section{Simulation details}

The details of the simulation are as in previous work~\cite{Ibuki2009} except
that either the Li$^+$-water interactions are explicitly modified, or
the geometry of the flexible water molecule is fixed, thus removing any
mechanical polarization that may exist, see e.g. figure~12 in~\cite{Ibuki2009}
for a distribution of the water molecular dipoles in LiCl solutions.
Consequently, two options were  pursued: a) to more or less arbitrarily reduce
the Coulomb part of the Li$^+$-oxygen and Li$^+$-hydrogen pair potentials and
b) to rigidify the water molecules in a reasonable geometry, see for comparison
figure~7 in reference~\cite{firstnacl} and, as mentioned, figure~12
in~\cite{Ibuki2009} for the distributions of angles and the molecular dipole
moments of flexible water in solutions.

%fig.2
\begin{figure}[ht]
\centerline{
\includegraphics[width=6.5cm,angle=-90]{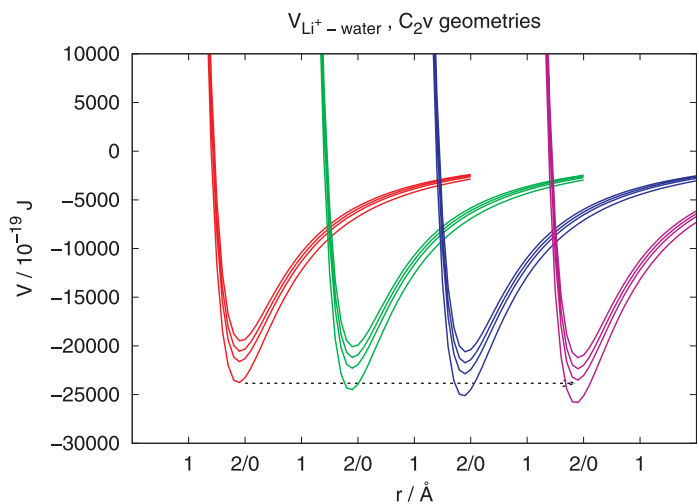}
}
\caption{(Color online) Li$^+$-water energies as a function of the Li$^+$--O distance $r$,
the parameter $\alpha$ ($\alpha =1.0 , 0.9 , 0.85 , 0.8$ from bottom to top),
and the geometry of the (flexible) water molecule
($r_{\rm OH} = 0.9572$~{\AA}, $\angle_{\rm HOH} = 104.5^\circ$, red curves;
$r_{\rm OH} = 0.9721$~{\AA}, $\angle_{\rm HOH} = 102.3^\circ$, green curves;
$r_{\rm OH} = 0.9872$~{\AA}, $\angle_{\rm HOH} = 101.7^\circ$, blue curves;
$r_{\rm OH} = 1.0115$~{\AA}, $\angle_{\rm HOH} = 100.0^\circ$, pink curves, the curves have been
shifted by 2~{\AA} in $r$-direction for better visibility.),
for C$_{2v}$ arrangements. The arrow shows that the potential at $\alpha=0.9$ for a
deformed water molecule corresponds to the full potential for water in its
approximate gas phase geometry.}
\label{pots}
\end{figure}

 In all other respects,  the simulations were identical to the previous ones:
556 BJH~\cite{bjh-intra} water molecules and, depending on concentration, 10 to
150 ion pairs at experimental densities and 300~K. We have simulated 1~m, 2~m,
4~m, 5~m, 8~m, 10~m, 12~m, and 15~m solutions. The simulations were run essentially
under ($NVE$)-conditions, except that a Kast-type thermostat~\cite{Kast1996}
was very loosely coupled (in order to minimize the perturbations of the
dynamics, see below) to compensate an unavoidable numerical error during the
$\approx$10$^6$ integration steps. Ewald summation was used throughout. Data
for the analysis were collected, after extensive equilibration runs, for
125~ps or 250~ps. The average temperature of the runs was  $ \langle T \rangle
= (298\pm$1.5)~K. In one test case we increased the box size by a factor of 27
(15012 water molecules and 2700 ion pairs, 10~m solution). Starting with the end
configuration of a smaller run we could extend this simulation only for about
6.75~ps. No changes in the structural data reported below could be detected in
this run.

The reduction in the cation-water interaction was simply achieved by
multiplying the Coulomb terms of the Li$^+$--O and Li$^+$--H pair potentials by
factors $\alpha < 1$, which is tantamount to multiplying for these interactions
the partial charges of Li$^+$, O and H by $\sqrt{\alpha}$. While this is
inconsistent, e.g., from a dielectric standpoint (e.g.; which charges should be
used to  compute the dielectric constant $\epsilon$?), this is sufficient for our
purposes here. We have applied factors of $0.8 \leqslant \alpha \leqslant 1.0$.
Figure~\ref{pots} shows the energies for a Li$^+$-water supermolecule with
C$_{2v}$ geometry obtained for various values  of $\alpha$ and for  rigidified
flexible BJH water molecules.  We have also conducted simulations with such
rigidified water (always with full charges, $\alpha=1$) using two slightly
different geometries: $r_{\rm OH} =0.9572$~{\AA}, $\angle_{\rm HOH}=
104.52^\circ$ and    $\angle_{\rm HOH} = 109.43^\circ$.

\section{Results and discussion}

\begin{figure}[!ht]
\centerline{
\includegraphics[width=4.7cm, angle=-90]{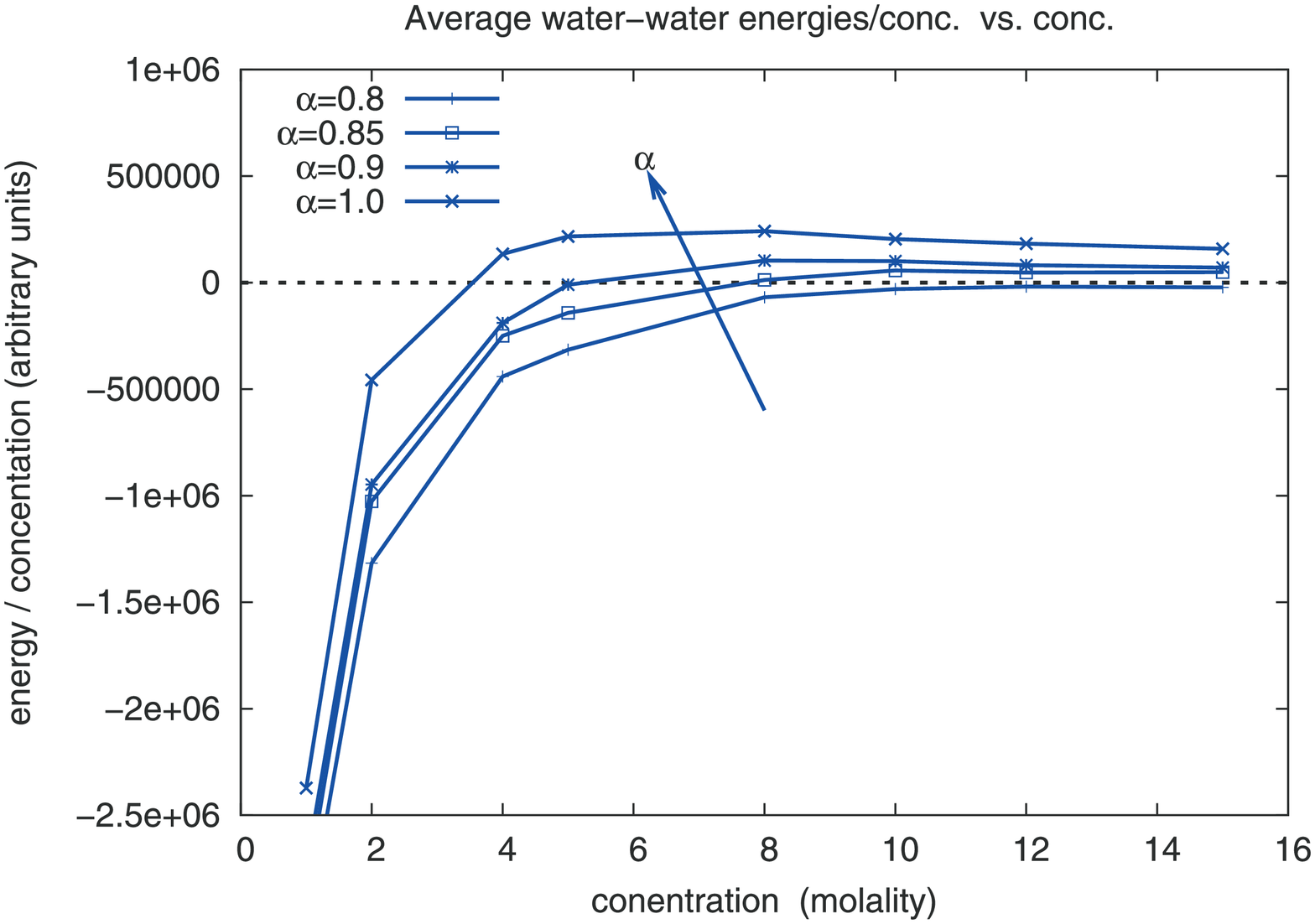}
\hfill
\includegraphics[width=4.7cm, angle=-90]{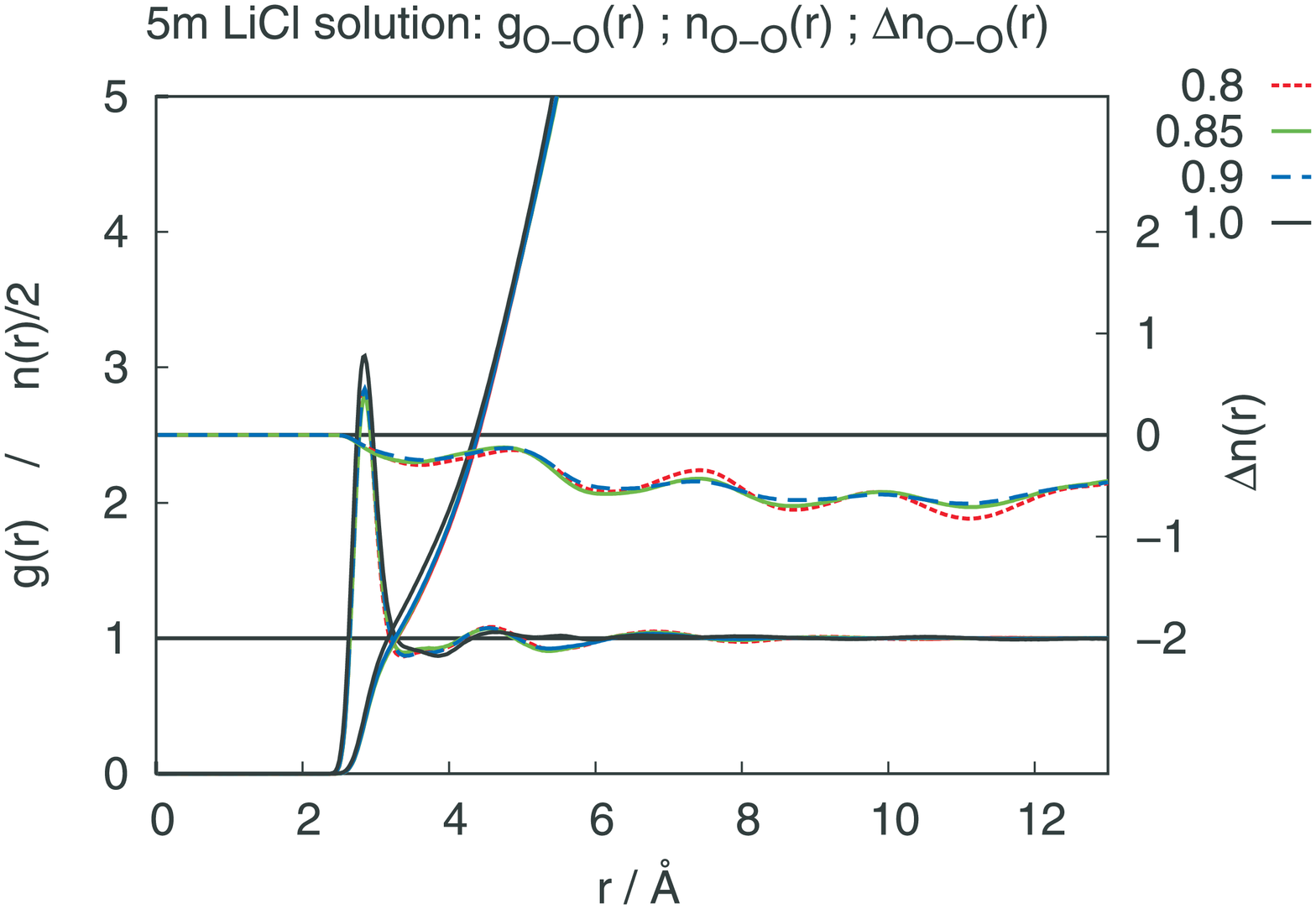}}
\vspace{2ex}
\centerline{
\includegraphics[width=4.7cm, angle=-90]{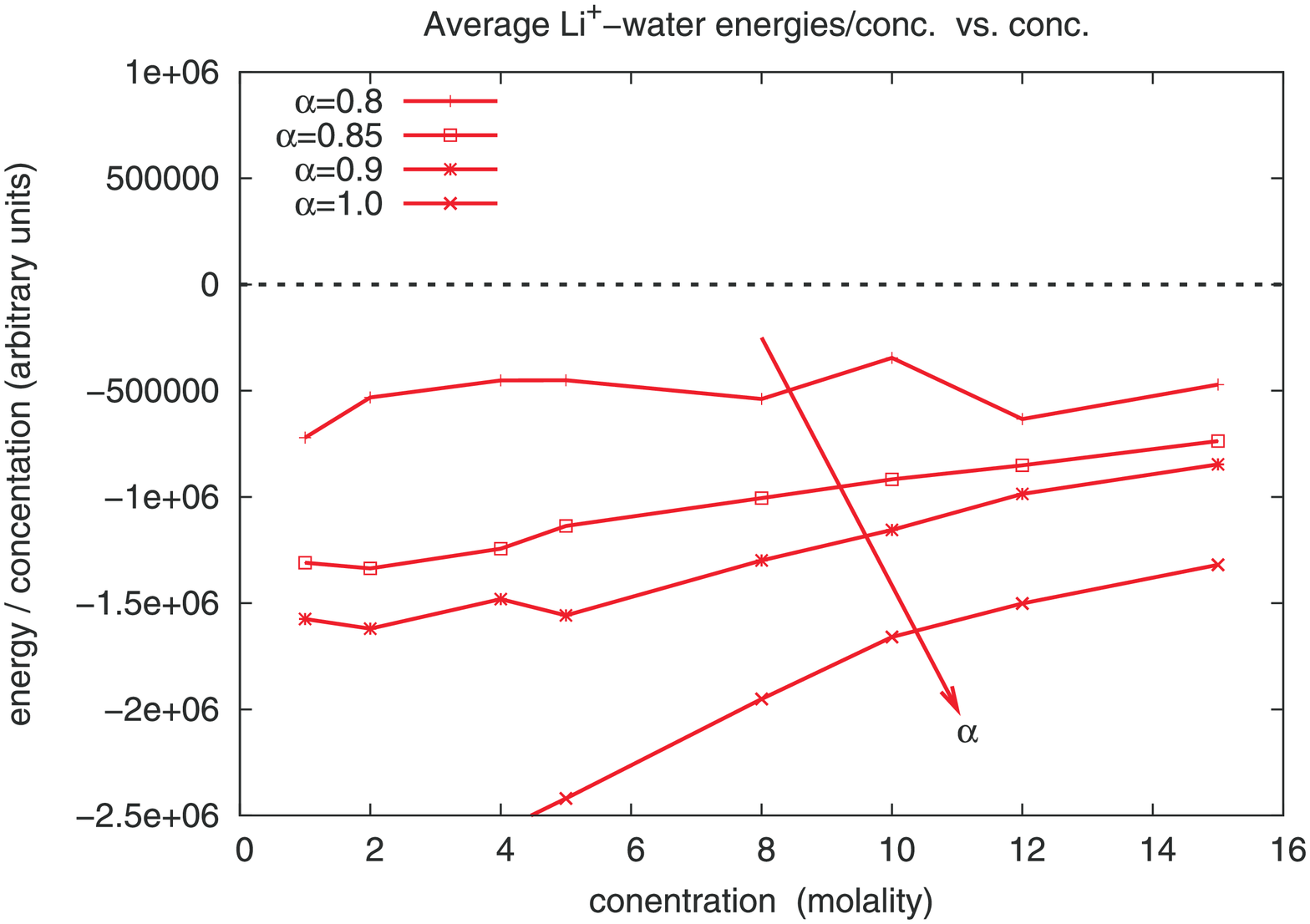}
\hfill
\includegraphics[width=4.7cm, angle=-90]{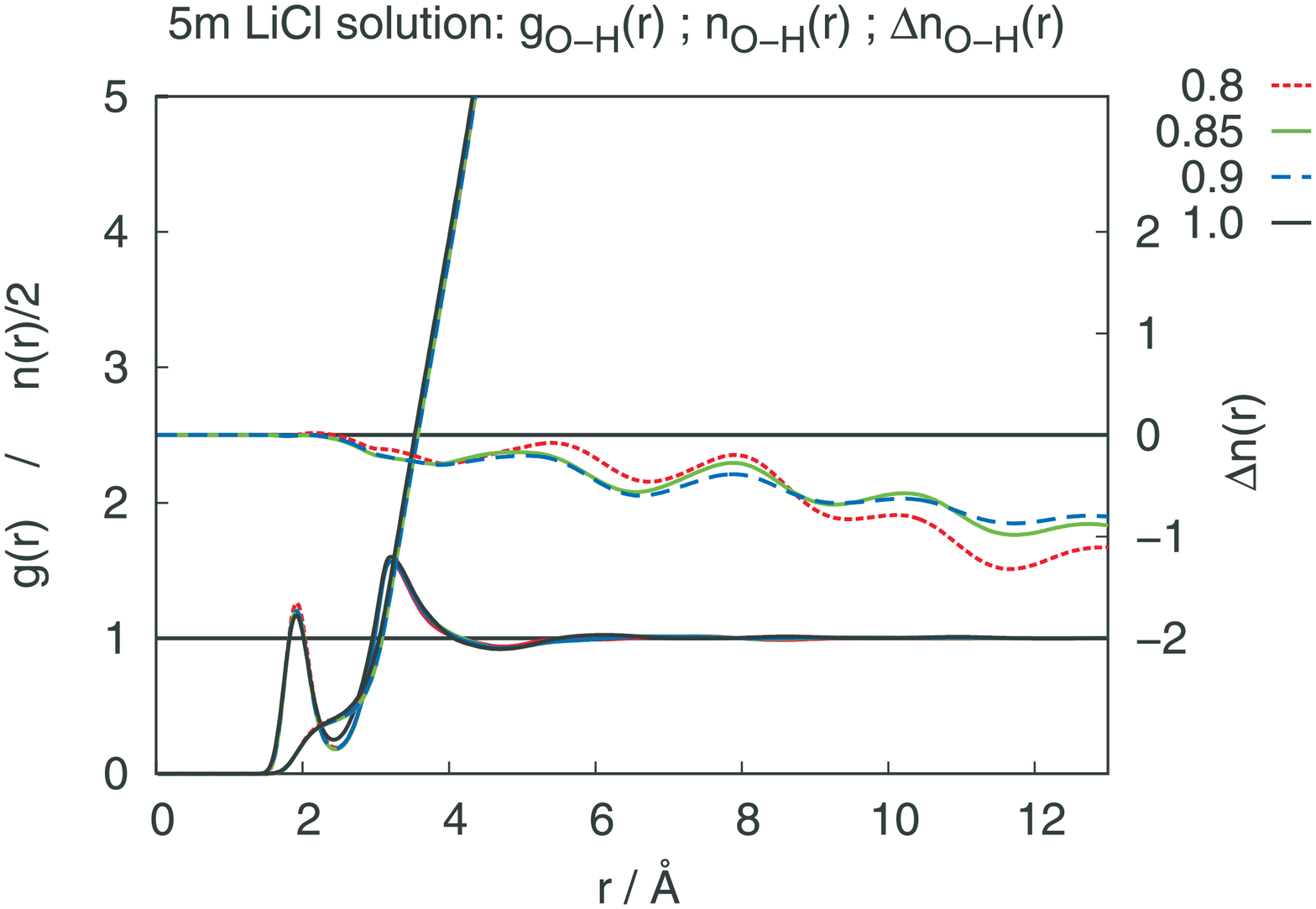}}
\vspace{2ex}
\centerline{
\includegraphics[width=4.7cm, angle=-90]{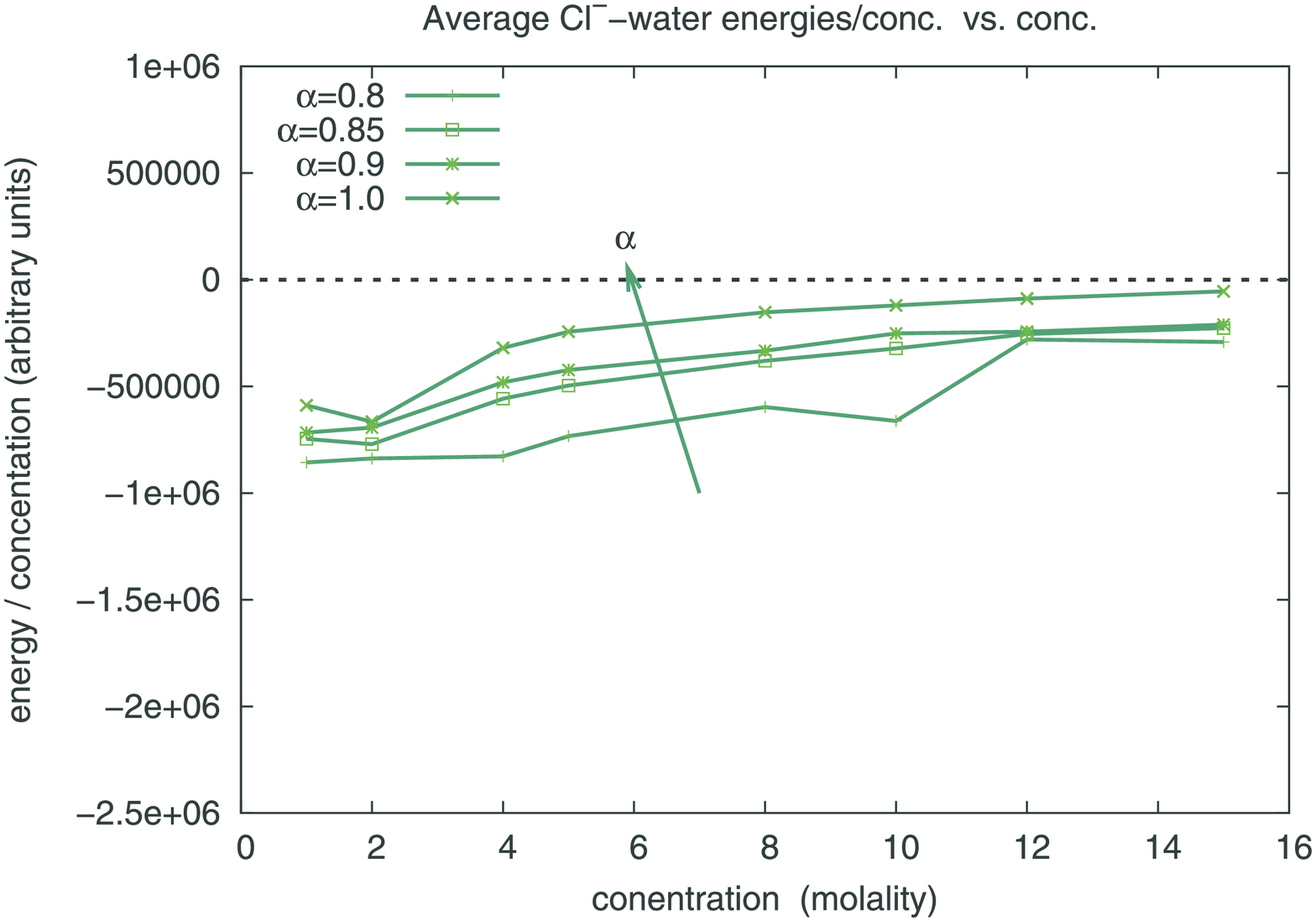}
\hfill
\includegraphics[width=4.7cm, angle=-90]{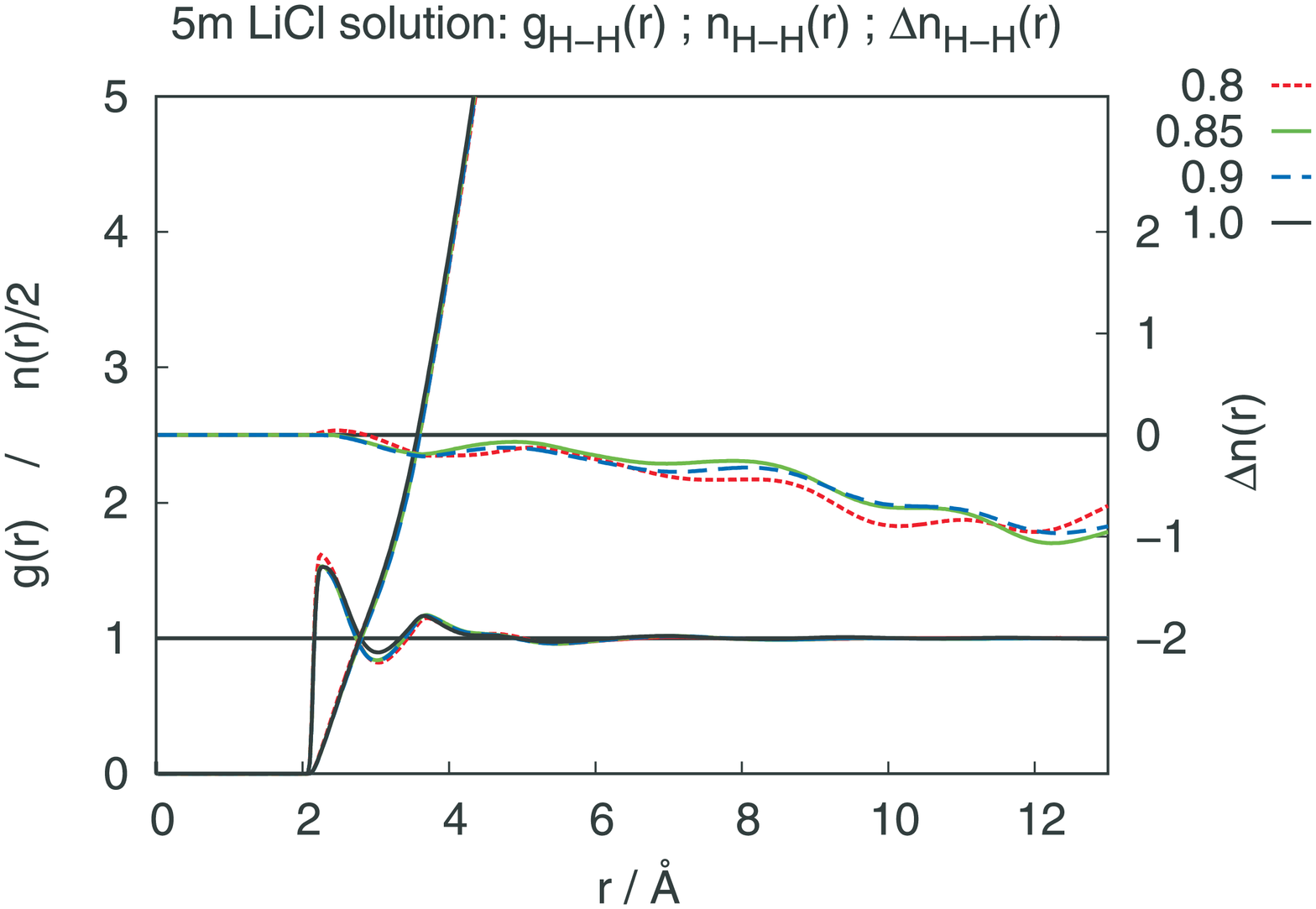}}
\centerline{
\parbox[t]{0.48\textwidth}{\caption{(Color online) Average water-water (top), Li$^+$-water (middle), and
Cl$^-$-water (bottom) energies, divided by the molality of the
LiCl-solution, as a function of the
molality, for flexible water and
values of $\alpha$ = 0.8, 0.85, 0.9, and 1.0. The arrows point
in the direction of increasing $\alpha$.}
\label{energies}
}
\hfill
\parbox[t]{0.48\textwidth}{\caption{(Color online) Water rdf's  ($g_{\rm O-O}$ (top),
$g_{\rm O-H}$ (middle), and
$g_{\rm H-H}$ (bottom)) and their running integration numbers $n(r)$
(left abscissa, the labels give the values of $g(r)$ and
$0.5 n(r)$) ;
the variations of $n(r)$ ($\Delta n(r)$, right abscissa, see text) for
the 5~m LiCl solutions. Flexible water, different $\alpha$-values.
}
\label{g-water}
}}
\end{figure}

Figure~\ref{energies} shows for the flexible models some of the contributions
to the total potential energies, divided by the salt concentration for better
comparison, as a function of concentration, for the flexible water simulations,
with $\alpha$ as a parameter. The middle panel shows how lowering $\alpha$,
i.e. diminishing the Coulomb interaction between the Li$^+$-ion and the water,
changes this energy component in the simulations as a function of
concentration. The top and bottom panels show how the average water-water and
Cl$^-$-water energies react to this modification of the Li$^+$-water
interaction. It is seen how lowering (increasing the magnitude) of the
cation-water terms `pushes up' the water-water energies, and that they can even
become positive at higher concentrations for the larger $\alpha$s, as already
noted previously~\cite{Ibuki2009}. This means that, loosely speaking, the
water  molecules are, on average, unfavorably oriented with respect to each
other (i.e. not adopting  hydrogen-bond-type configurations), their
orientations being presumably controlled by the cations.  The  Cl$^-$-water
energies (with unmodified pair potentials)  increase in magnitude (i.e. are
lowered) when the Li$^+$-water terms are diminished. Other energy terms are
modified in a similar way, which will not be discussed in detail here.

%figs.3-4
%%%%%%%%%%%%%%%%%%%%%%%%%%%%%%%%%%%%%%%%%%%%%%%%%%%%
%%%%%%%%%%%%%%%%%%%%%%%%%%%%%%%%%%%%%%%%%%%%%%%%%%%

Figure~\ref{g-water} shows, as an example, the influence of $\alpha$
on the water (O--O, O--H, and H--H radial pair distribution functions
(rdf)) for the 5~m solutions. The plot shows the rdf's and the
`running integration number' (number of neighbors)  $n(r)$ (both left ordinate).
The modifications of these $g$-functions due to the modification of
the Li$^+$-water interaction are small, but clearly identifiable:
The number of neighbors becomes slightly less when $\alpha$ goes from
1 to 0.8, although the effect seems not to be linear.

To highlight the effect of the
changes in the Li$^+$--O potential on all $g$-functions in the system,
we have found it most instructive to plot the quantities
\begin{eqnarray*}
n(r) & = & 4  \pi  \rho \int_0^r g\left(r'\right)  {r'}^2 \ {\rd}r',\\
 \Delta n(r) = n^{\rm rc}(r) - n^{\rm fc}(r) & = &
4  \pi  \rho \int_0^r \left[g^{\rm rc}\left(r'\right) - g^{\rm fc}\left(r'\right)\right]
 {r'}^2 \ {\rd}r',
\end{eqnarray*}
where rc and fc stand for `reduced charges' ($\alpha$=0.8, 0.85, 0.9)
and `full charges' ($\alpha =1$) in the Li$^+$-water interactions,
respectively. By analogy, rc will also be used for the
rdf's obtained from rigid water simulations.
Figure~\ref{g-water} also shows $\Delta n(r)$ (right ordinate).
The overall effect of $\alpha$ and concentration on $\Delta n(r)$
is studied in a more systematic way in figures~\ref{3dg1}.

\begin{figure}[ht]
\centerline{
\includegraphics[width=5cm, angle=-90]{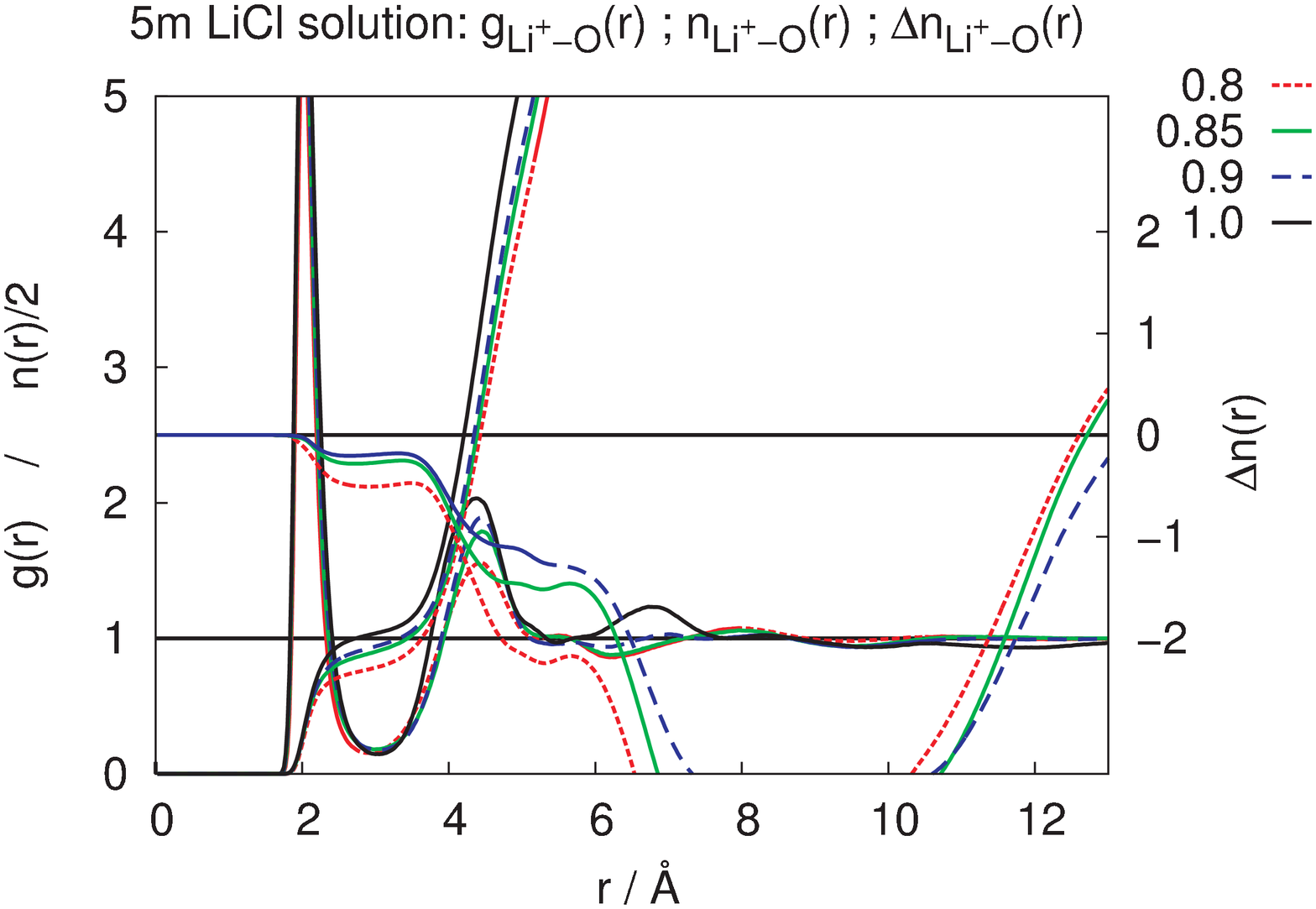}
\hfill
\includegraphics[width=5cm, angle=-90]{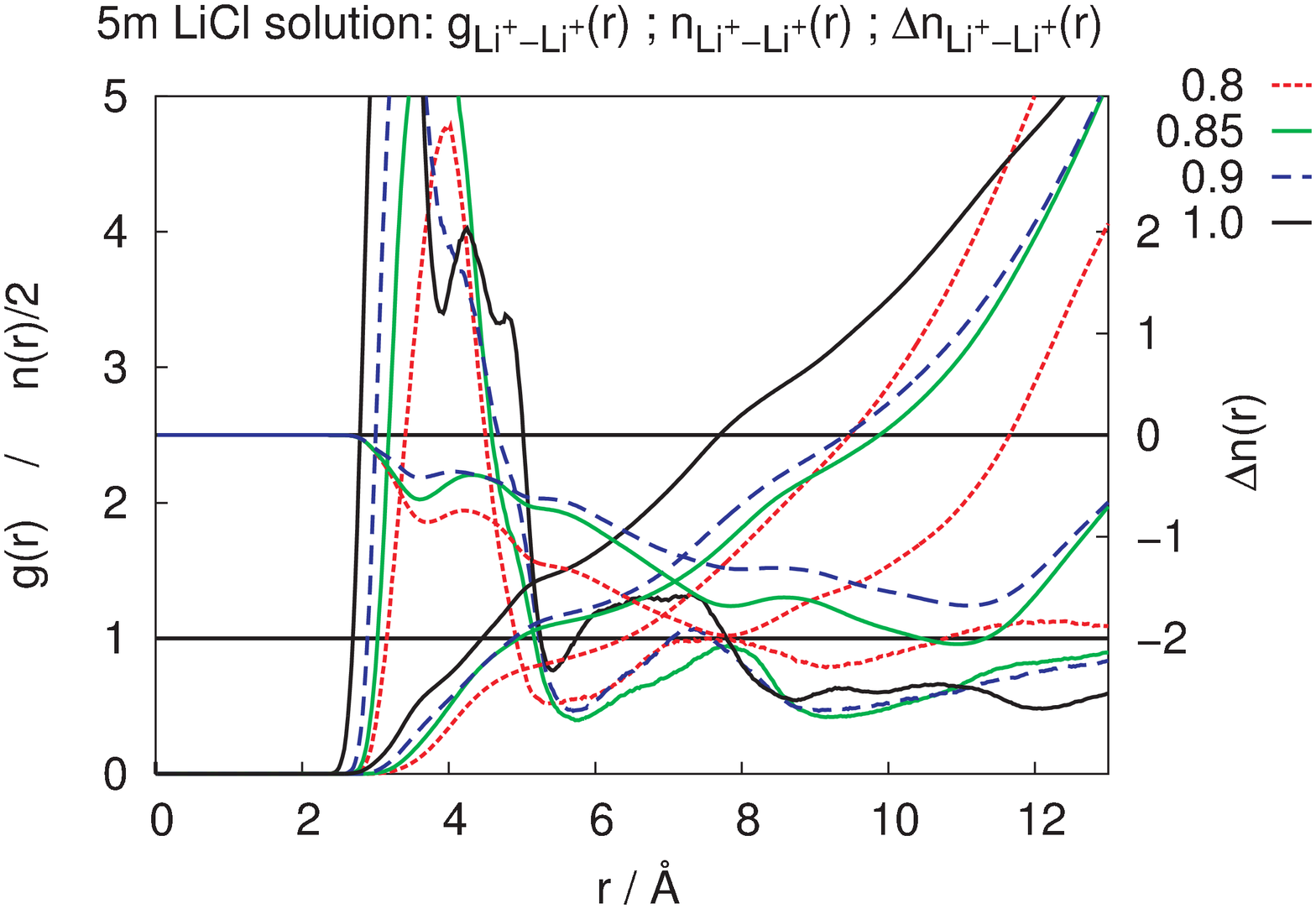}
}
\caption{(Color online) $g_{\rm Li^+-O}$ and  $g_{\rm Li^+-Li^+}$ rdf's,
running integration numbers and $\Delta n(r)$
for the 5~m LiCl solutions,
same representation as figure~\ref{g-water}. The curves off-scale here
(to keep the same axes as in figure~\ref{g-water}) are seen fully
in figure~\ref{3dg1}.}
\label{g-ii}
\end{figure}

\begin{figure}[!ht]
\centerline{
\includegraphics[width=3.6cm,angle=-90]{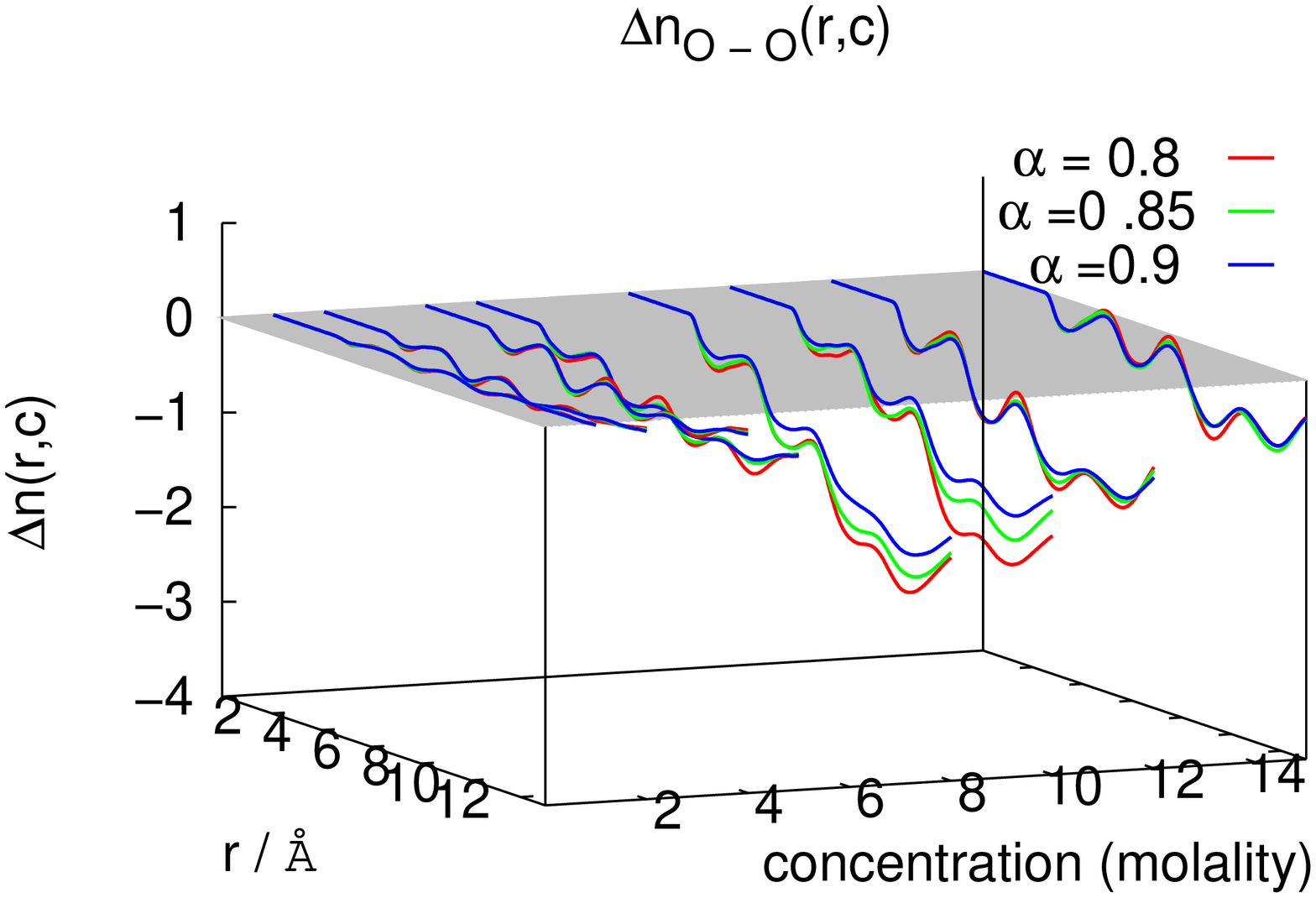}
\hspace{-0.6cm}
\includegraphics[width=3.6cm,angle=-90]{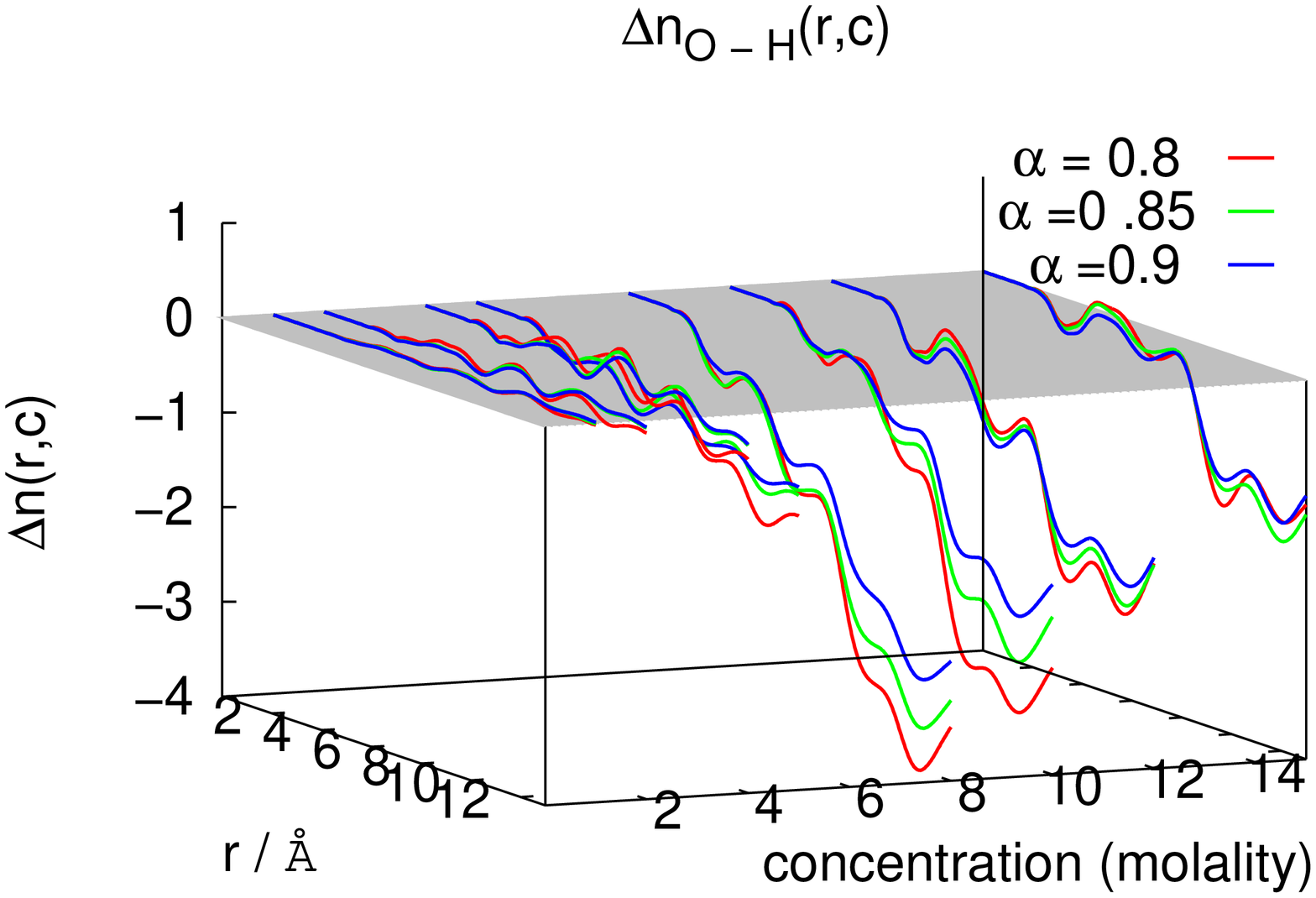}
\hspace{-0.6cm}
\includegraphics[width=3.6cm,angle=-90]{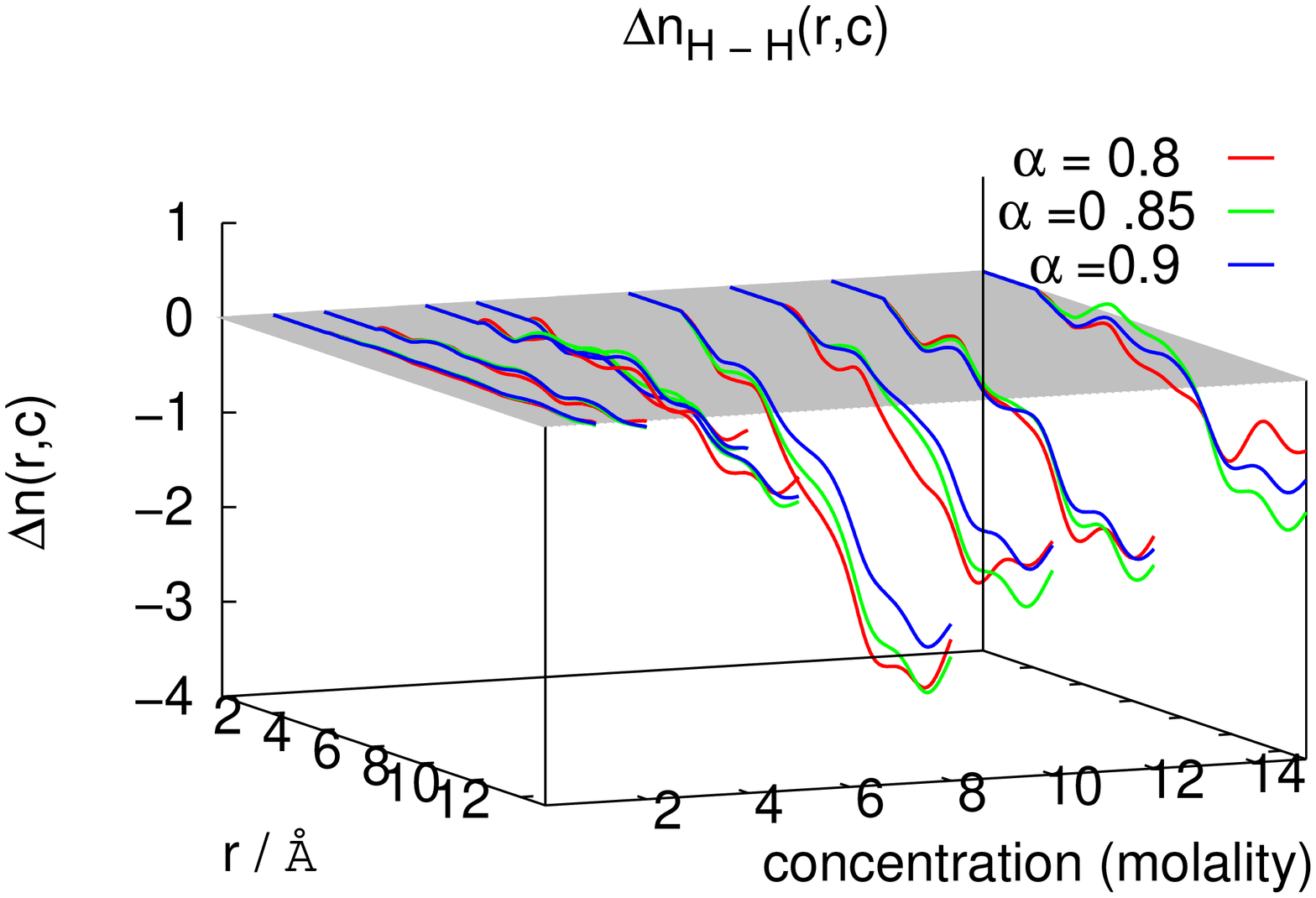}
\hspace{-0.2cm}
\raisebox{-2cm}{\ (a) \/}
}
\vspace{-3mm}

\begin{flushright}
\includegraphics[width=3.6cm,angle=-90]{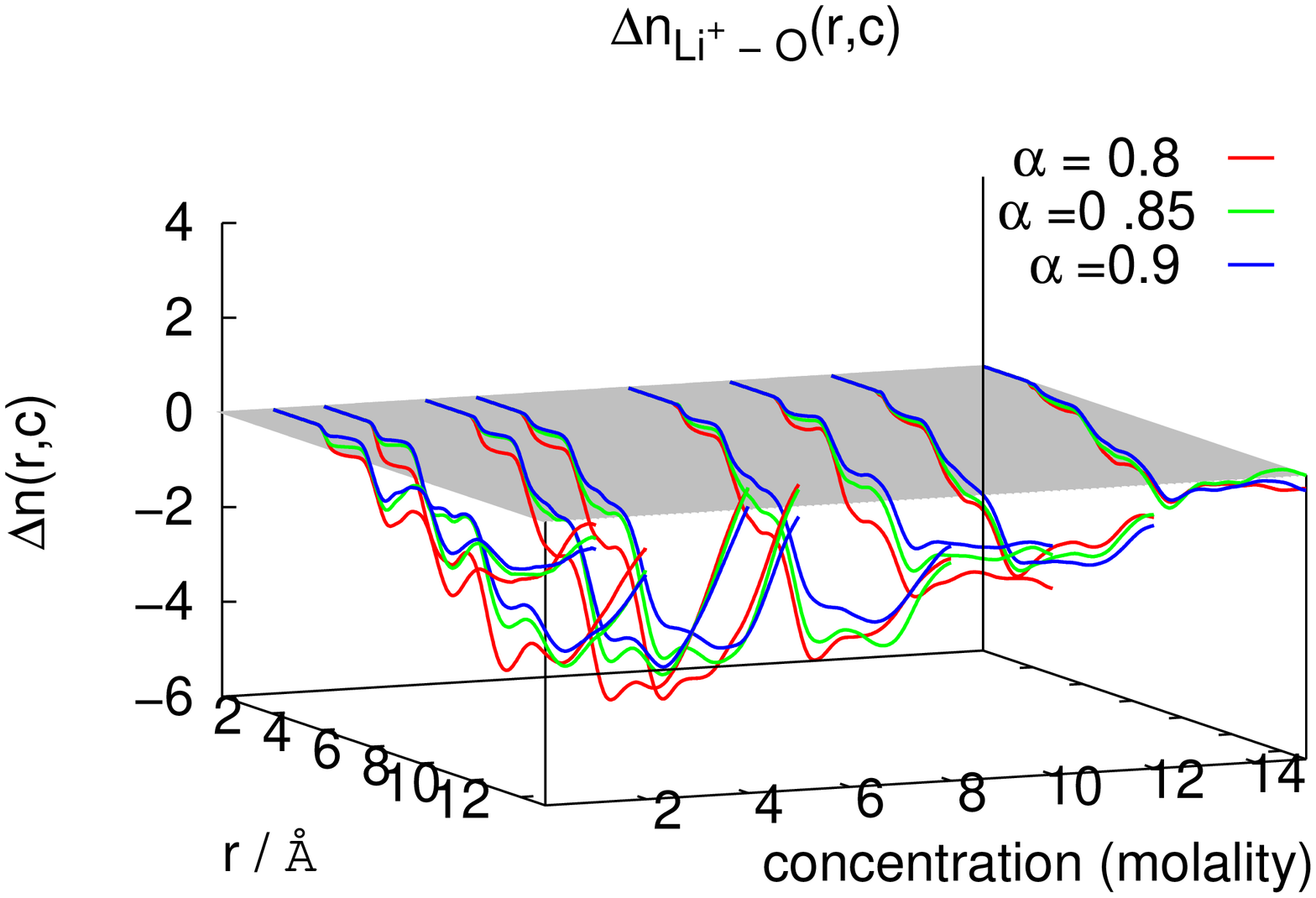}
\hspace{-0.6cm}
\includegraphics[width=3.6cm,angle=-90]{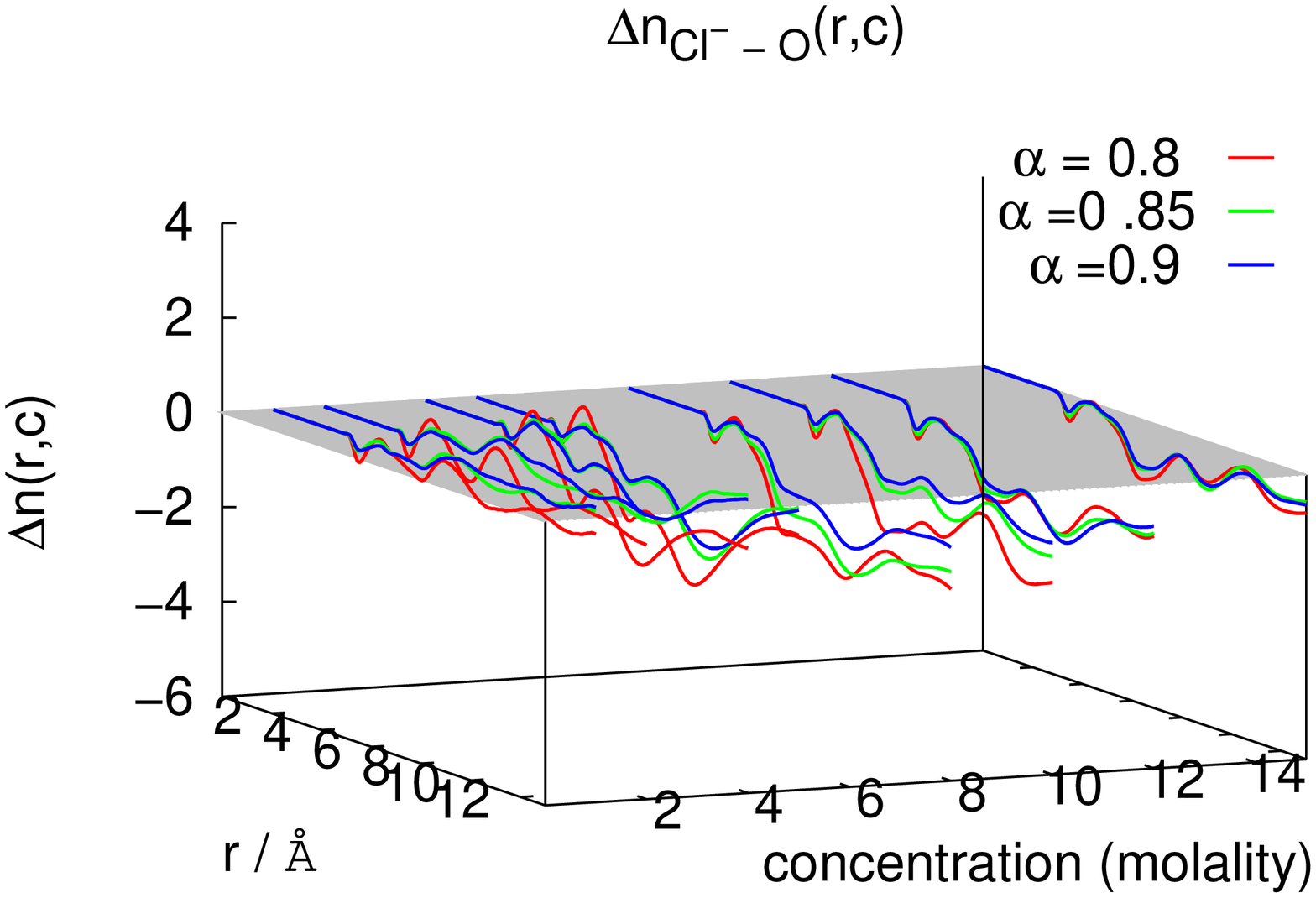}
\hspace{1.6cm}
\raisebox{-2cm}{\ (b) \/}
\end{flushright}
\vspace{-6mm}

\begin{flushright}
\includegraphics[width=3.6cm,angle=-90]{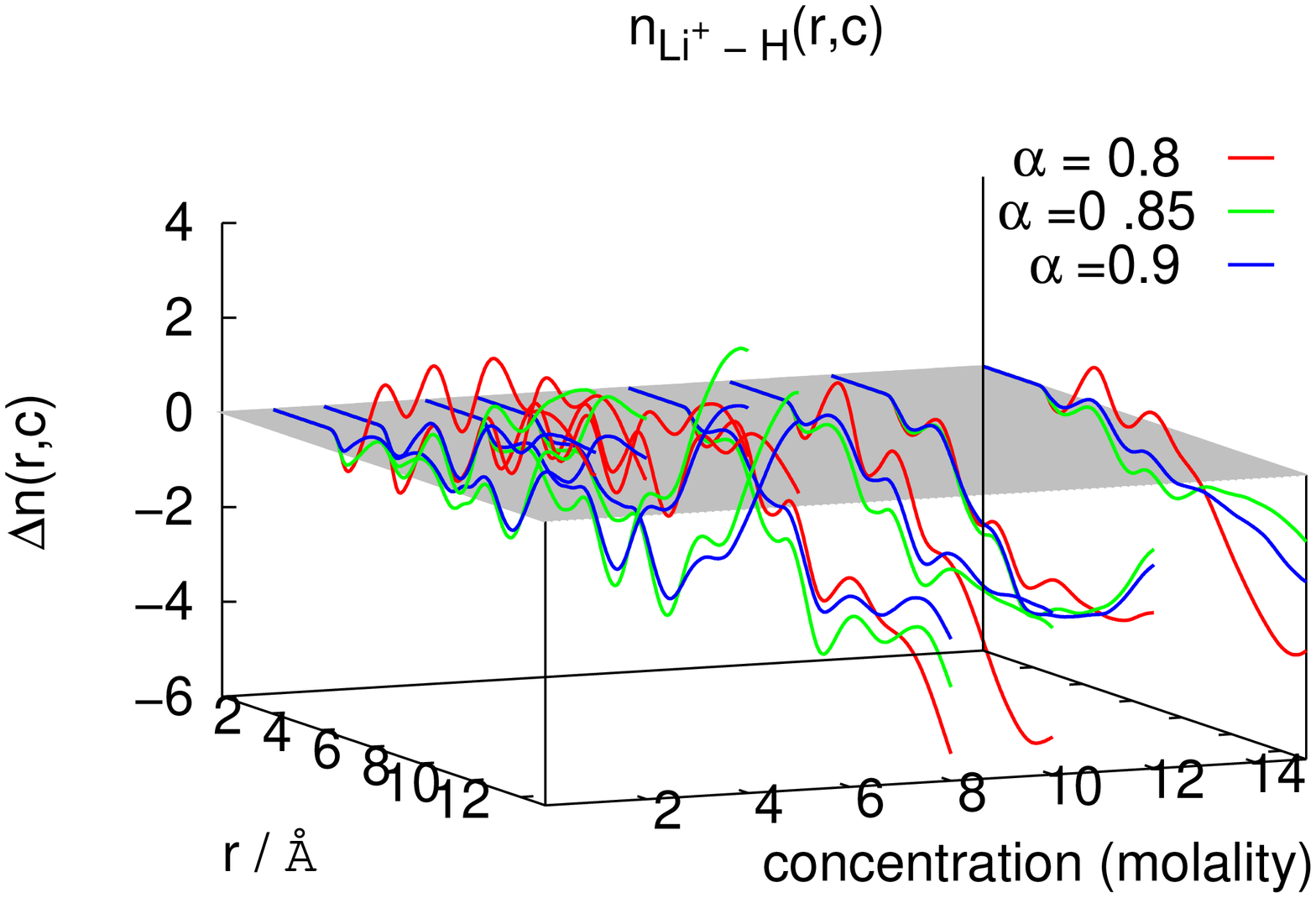}
\hspace{-0.6cm}
\includegraphics[width=3.6cm,angle=-90]{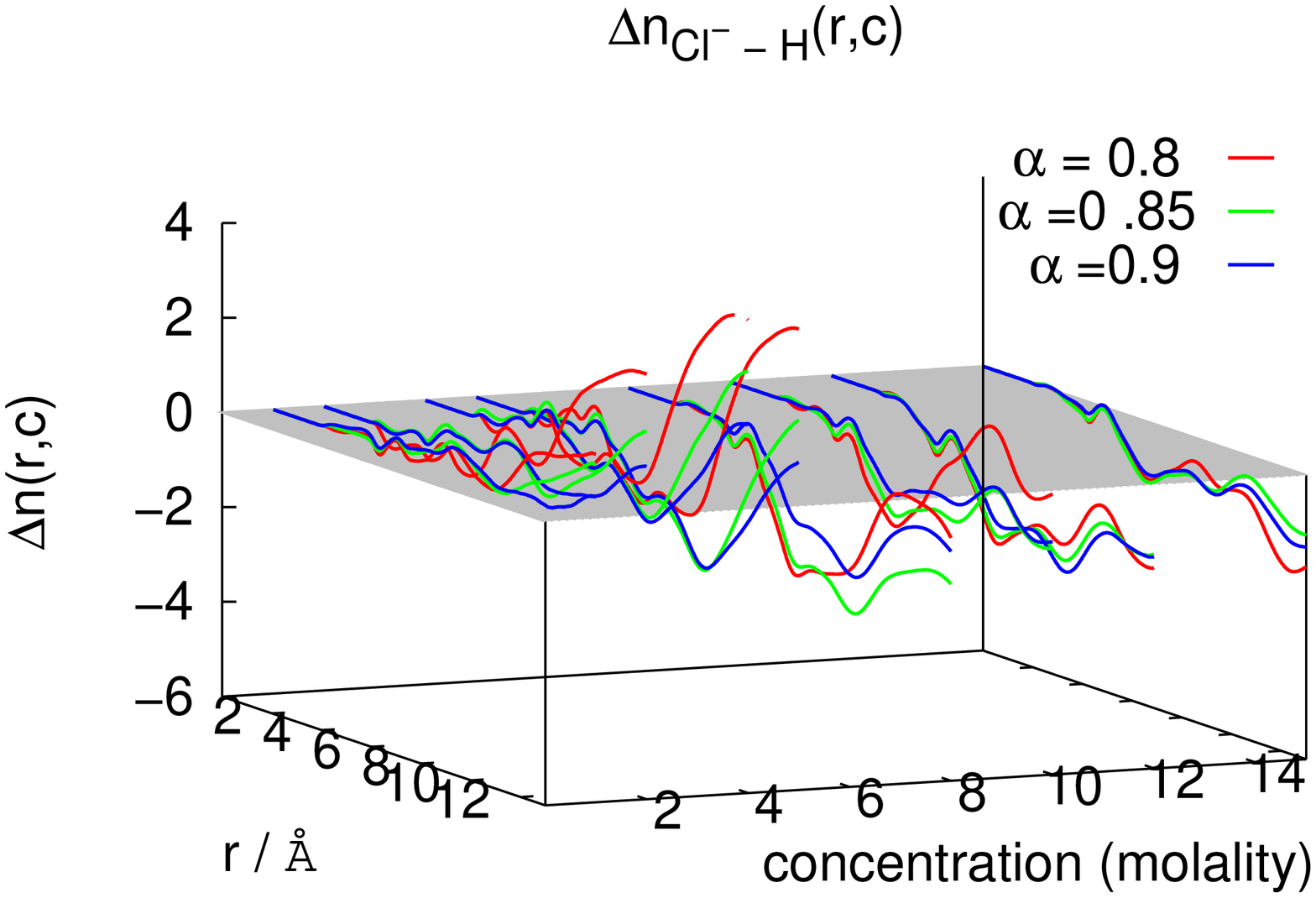}
\hspace{1.6cm}
\raisebox{-2cm}{\ (c) \/}
\end{flushright}
\vspace{-3mm}

\centerline{
\includegraphics[width=3.6cm,angle=-90]{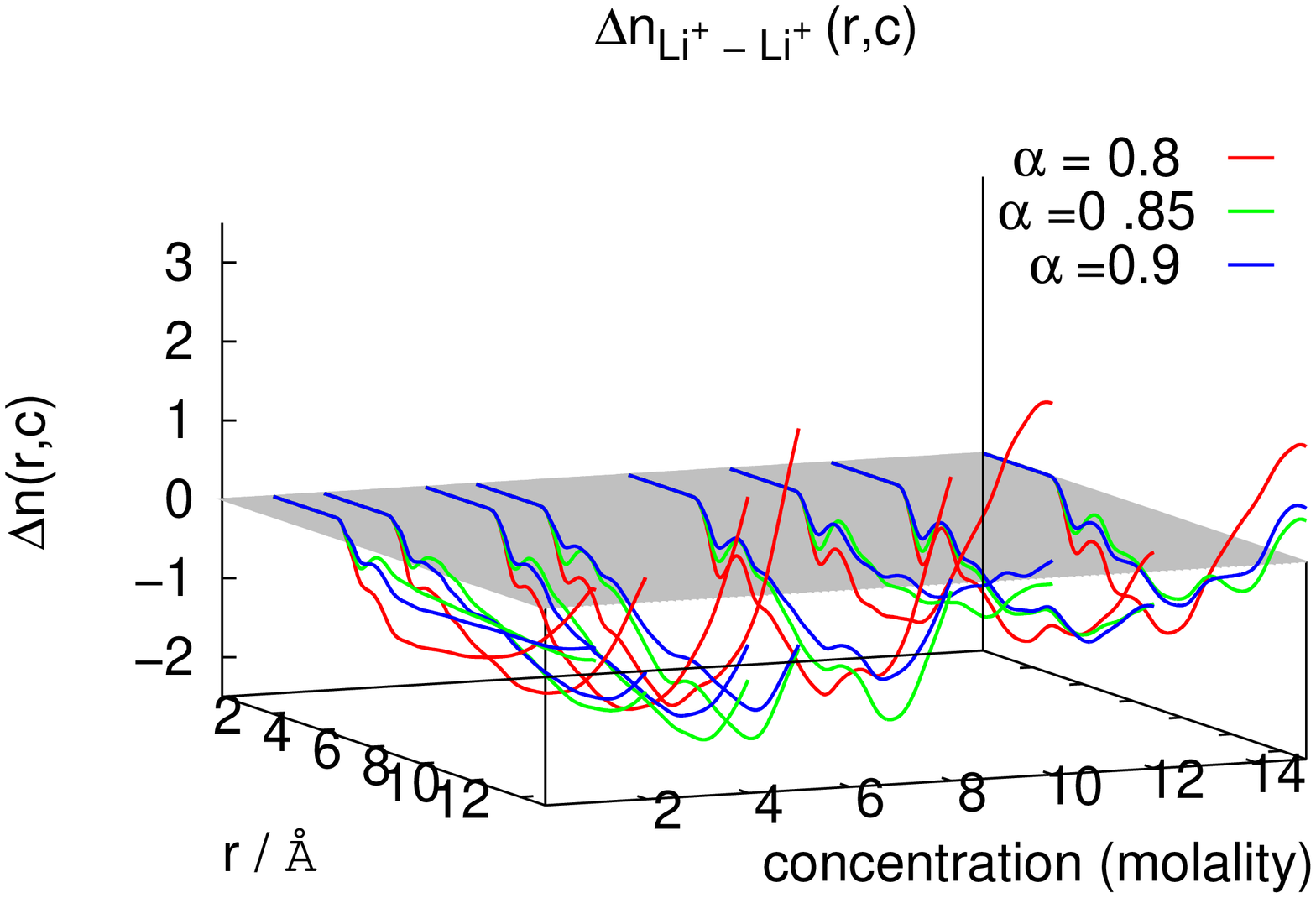}
\hspace{-0.6cm}
\includegraphics[width=3.6cm,angle=-90]{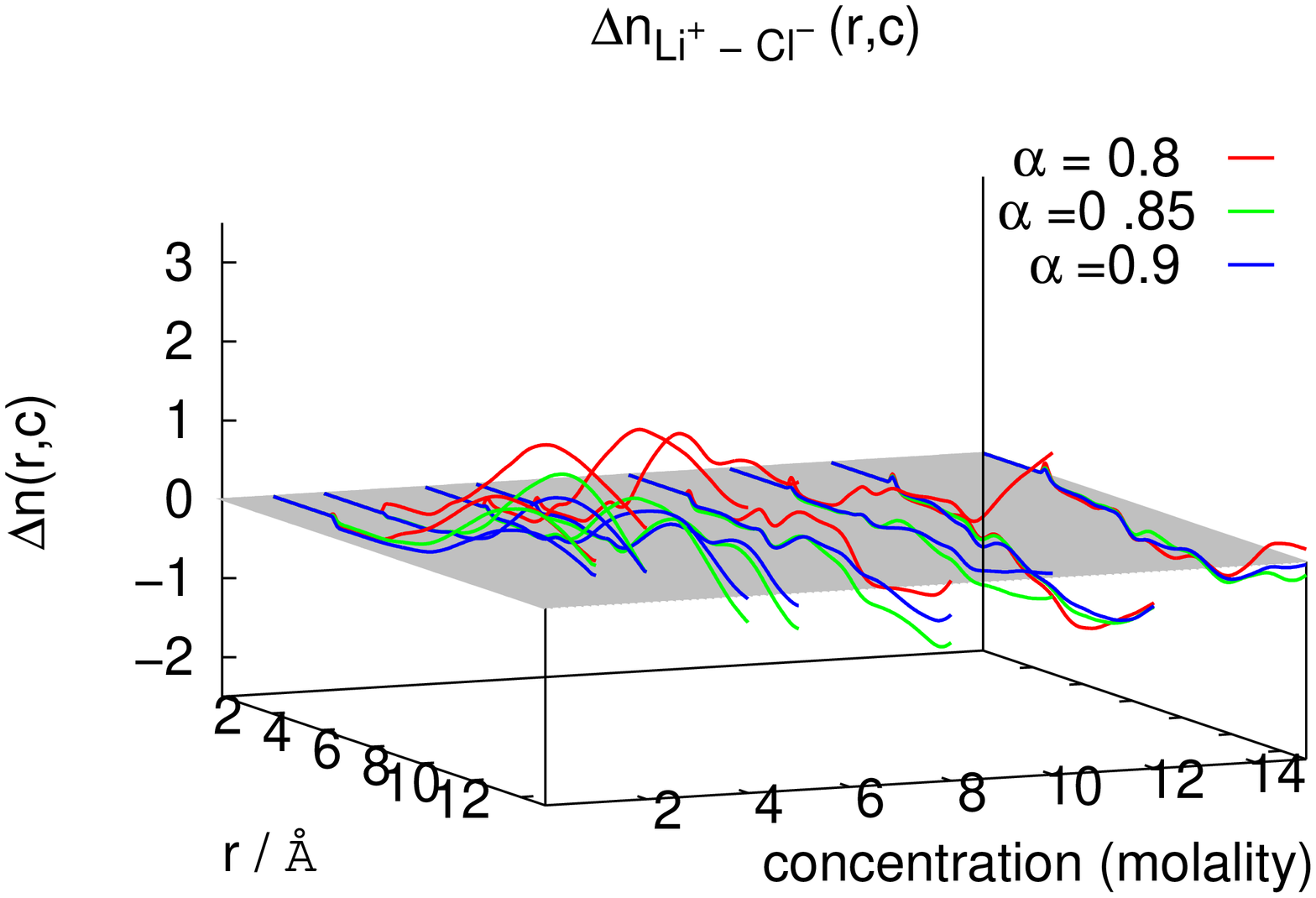}
\hspace{-0.6cm}
\includegraphics[width=3.6cm,angle=-90]{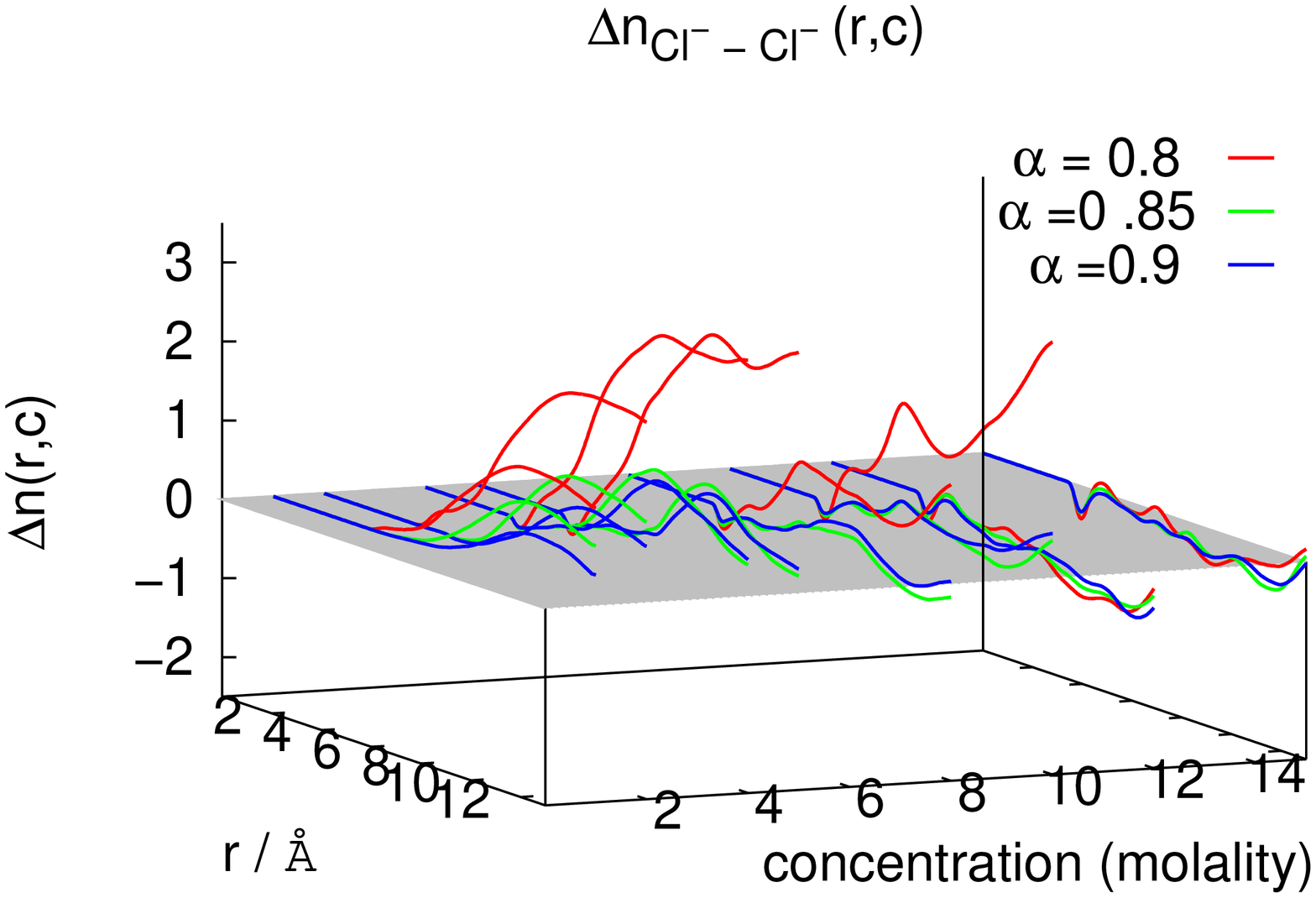}
\hspace{-0.3cm}
\raisebox{-2cm}{\ (d) \/}
}

\caption{(Color online) Variation of the running integration number as a function of
$r$ and salt concentration,
($\Delta n(r,{\rm concentration})$ , see text),
for flexible water simulations and different values of $\alpha$.
The grey plane is the reference: $\alpha =1$, where $\Delta n(r)=0$.
Panel (a):
O--O, O--H, and H--H functions (water-water), panels (b), (c):
Li$^+$--O, Li$^+$--H, Cl$^-$--O, and Cl$^-$--H functions (ion-water),
panel (d): Li$^+$--Li$^+$, Li$^+$--Cl$^-$, and Cl$^-$--Cl$^-$ functions
(ion-ion). Note the different scales for the ordinate in different rows.
}
\label{3dg1}
\end{figure}

Figure~\ref{g-ii} is analogous to figure~\ref{g-water}
for the Li$^+$--O and Li$^+$--Li$^+$
functions. The effect of changing $\alpha$ is here much more pronounced
and goes in the same direction as observed above for the water
functions. It is seen in particular that the Li$^+$--Li$^+$ aggregation
is loosened: For $\alpha=0.8$, the number of next Li$^+$-neighbors of a central
Li$^+$-ion is almost halved up to distances of about
8~{\AA}.

%fig.5

Figure~\ref{3dg1} shows $\Delta n(r)$ as a function of
$r$ and the concentration for the three reduced charges.
The panel~(a) of this figure refers to the three $g$-functions of water,
the panels~(b) and (c)  to the ion-water and the panel~(d) to the ion-ion
functions.  Inspection of this figure and figures~\ref{g-water} and~\ref{g-ii}  reveals that modifying just the Li$^+$-water interaction has a detectable  effect on all radial pair distributions. This effect of
$\alpha$  is generally most pronounced at intermediate concentrations (between
about 6 and 12~m, say). As noted, the number of neighbors, i.e., the local
particle density, is generally, at least up to intermediate distances,
decreased when $\alpha$ goes from 1 to 0.8. The Cl$^-$--Cl$^-$ functions are an
exception. We note here that the overall system densities were the same
for all $\alpha$s (($NVE$)-simulations).

\begin{figure}[!ht]
\centerline{
\includegraphics[width=4.8cm, angle=-90]{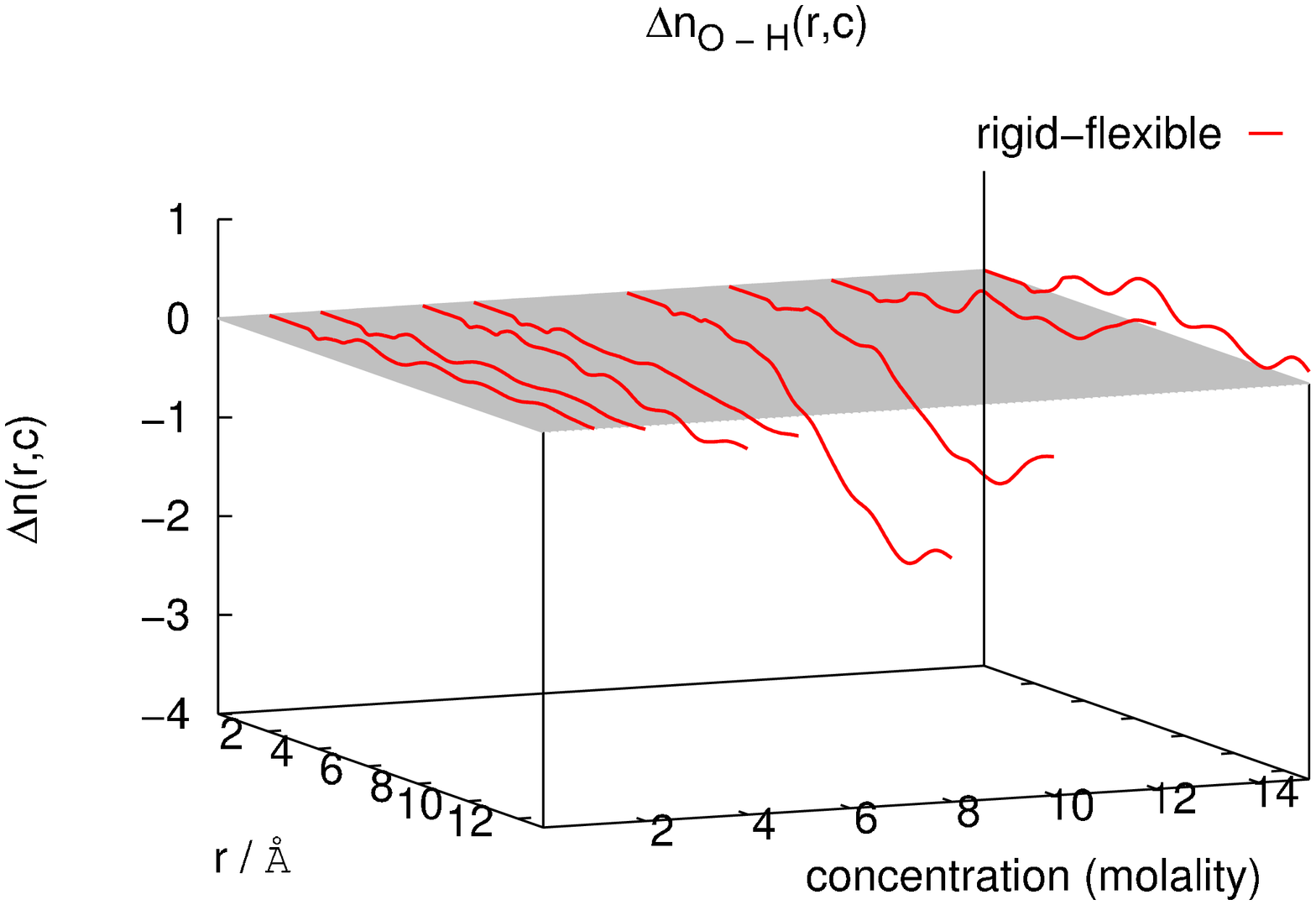}
\hfill
\includegraphics[width=4.8cm, angle=-90]{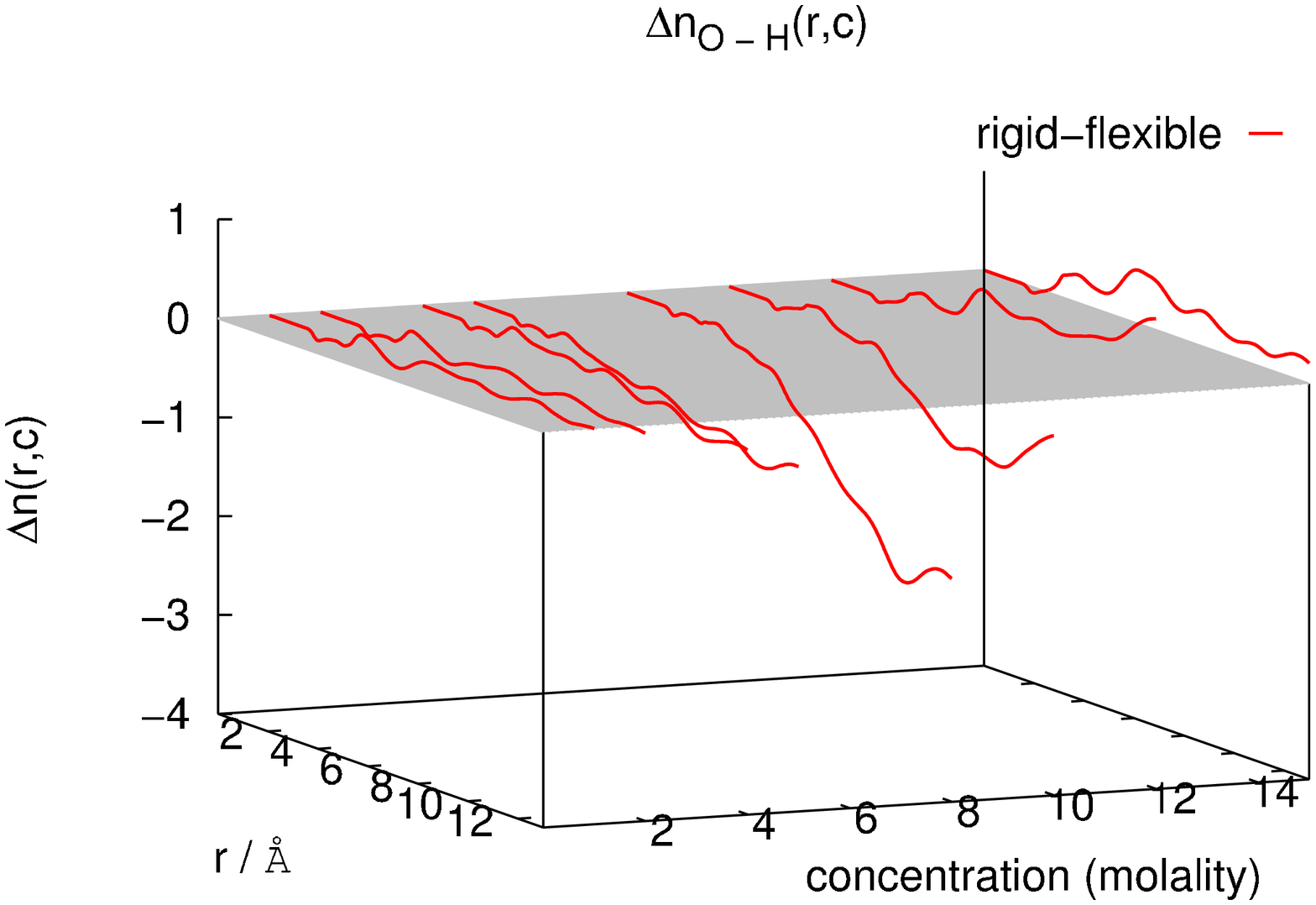}
}
\vspace{5mm}
\centerline{
\includegraphics[width=4.8cm, angle=-90]{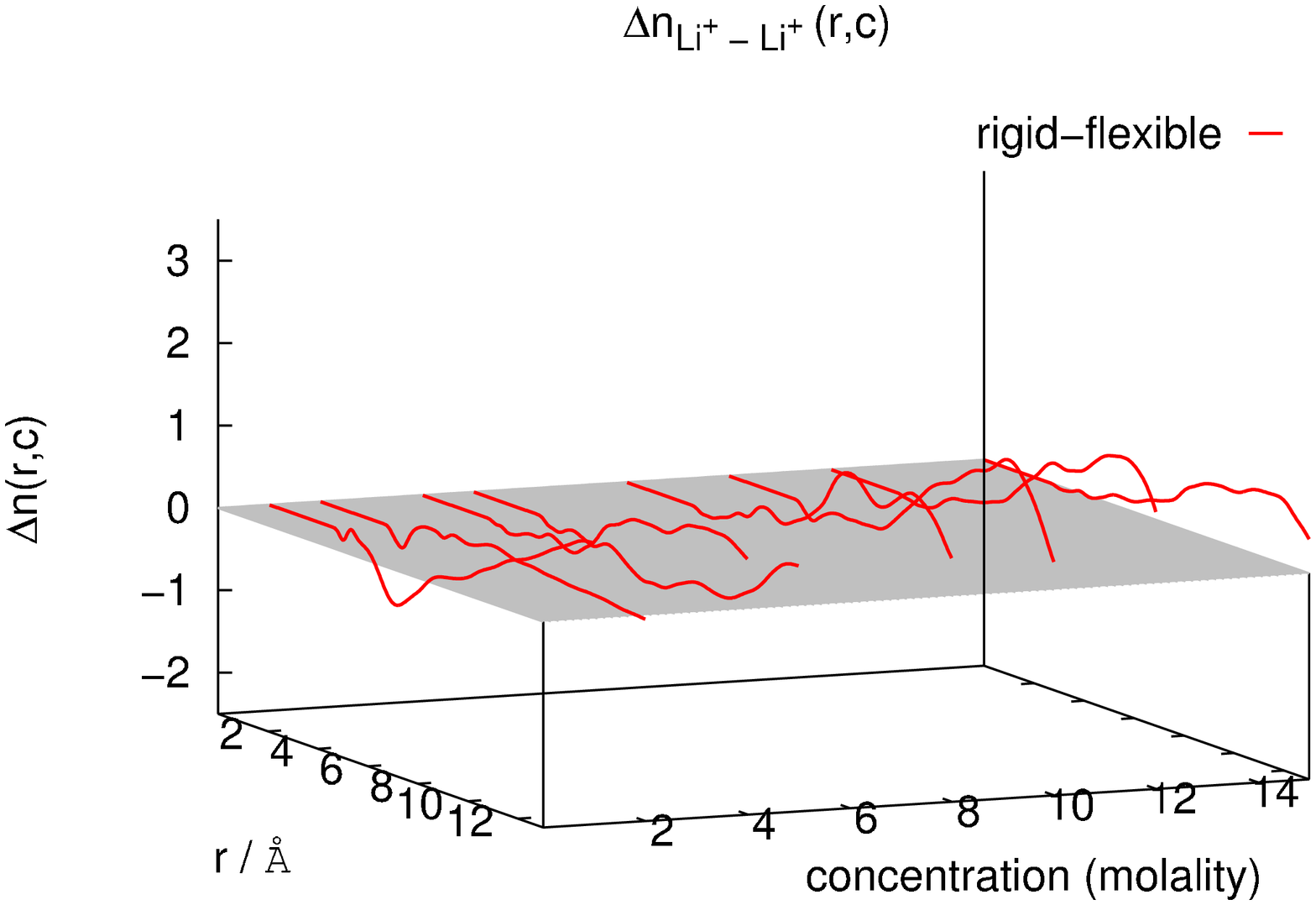}
\hfill
\includegraphics[width=4.8cm, angle=-90]{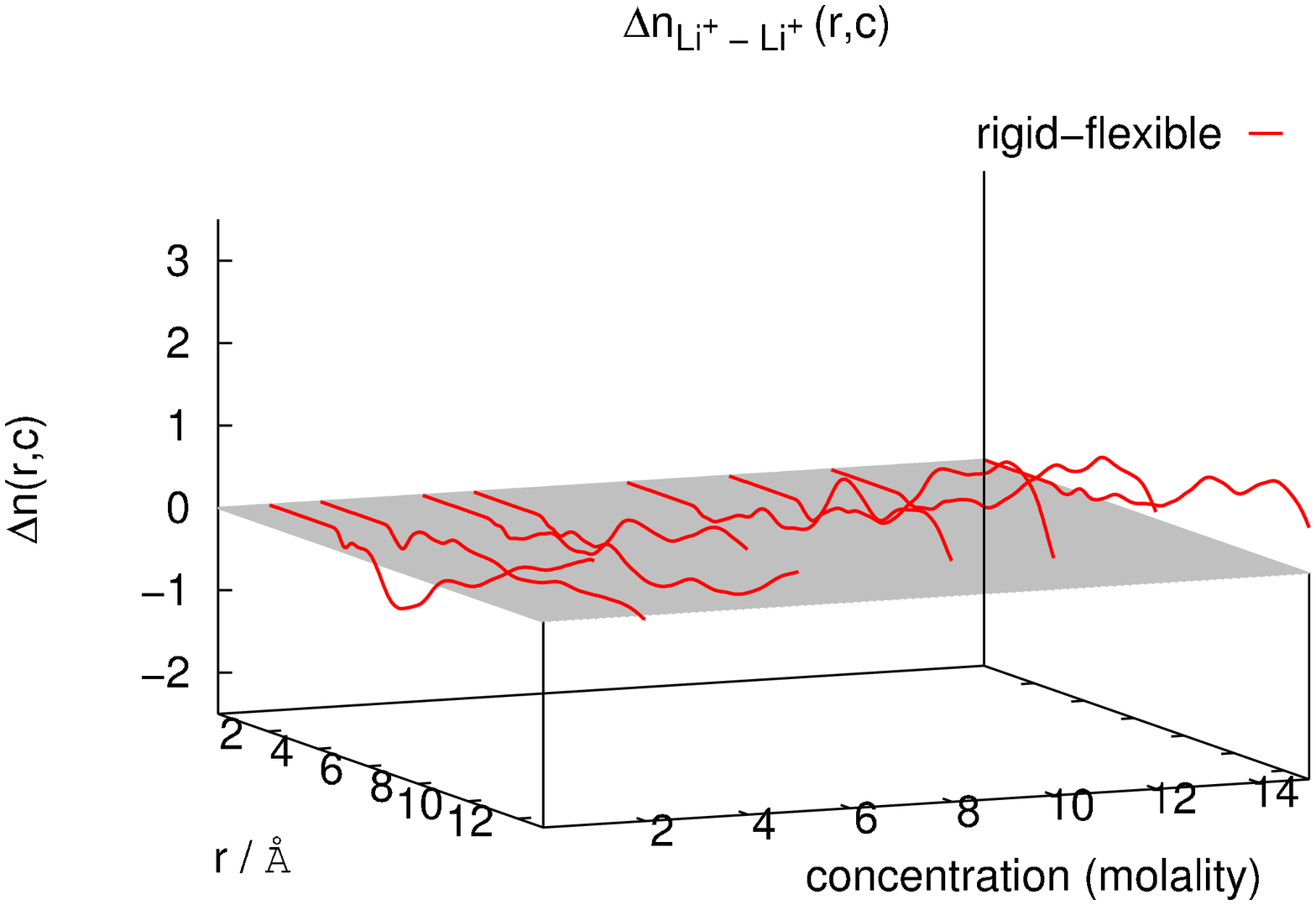}
}
\caption{(Color online) Selected panels like in figure~\ref{3dg1}, but for simulations with
 rigid water molecules:   $r_{\rm OH} =0.9572$~{\AA},
$\angle_{\rm HOH}= 104.52^\circ$ (left) and
$\angle_{\rm HOH} = 109.43^\circ$ (right).
}
\label{3dg2}
\end{figure}

Figure~\ref{3dg2} shows that rigidifying the water molecules, i.e., reducing
their average intermolecular energies (due to the absence of the
mechanical polarization of the flexible model) leads to similar, but
smaller changes in the Li$^+$--Li$^+$ rdf as explicitly reducing
the Li$^+$-water interaction. Similar observations
are made for the other $g$-functions.

%fig.7

%fig.8
\begin{figure}[!b]
\centerline{
\includegraphics[width=0.48\textwidth]{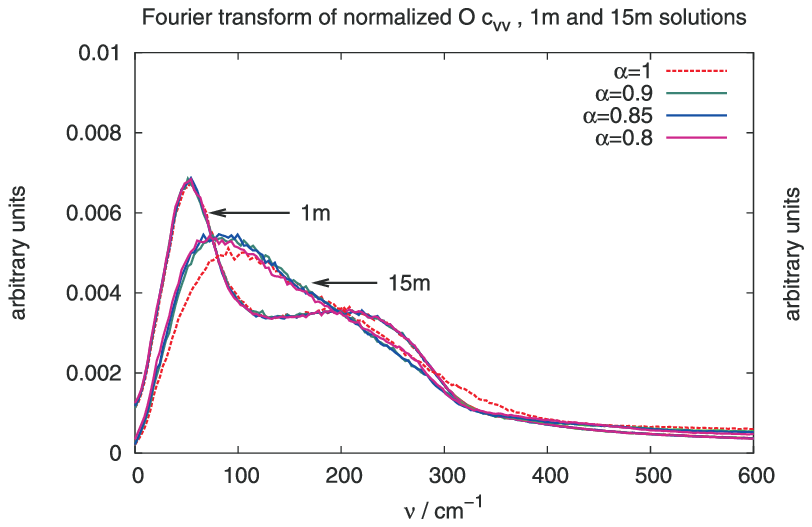}
\hfill
\includegraphics[width=0.48\textwidth]{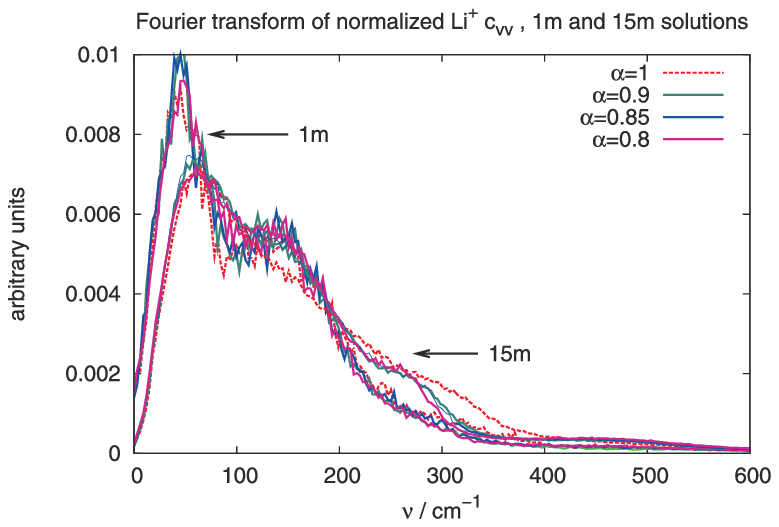}
}
\caption{(Color online) Spectral densities of motion (Fourier-cosine transforms
of the velocity autocorrelation function) for the oxygen atoms
of the water molecules (left) and the Li$^+$-ions (right)
in the 1 molal and 15 molal solutions for flexible water and
the 4 different values of $\alpha$.}
\label{ftcvv}
\end{figure}

%fig.9
\begin{figure}[!ht]
\centerline{
\includegraphics[width=6.5cm,angle=-90]{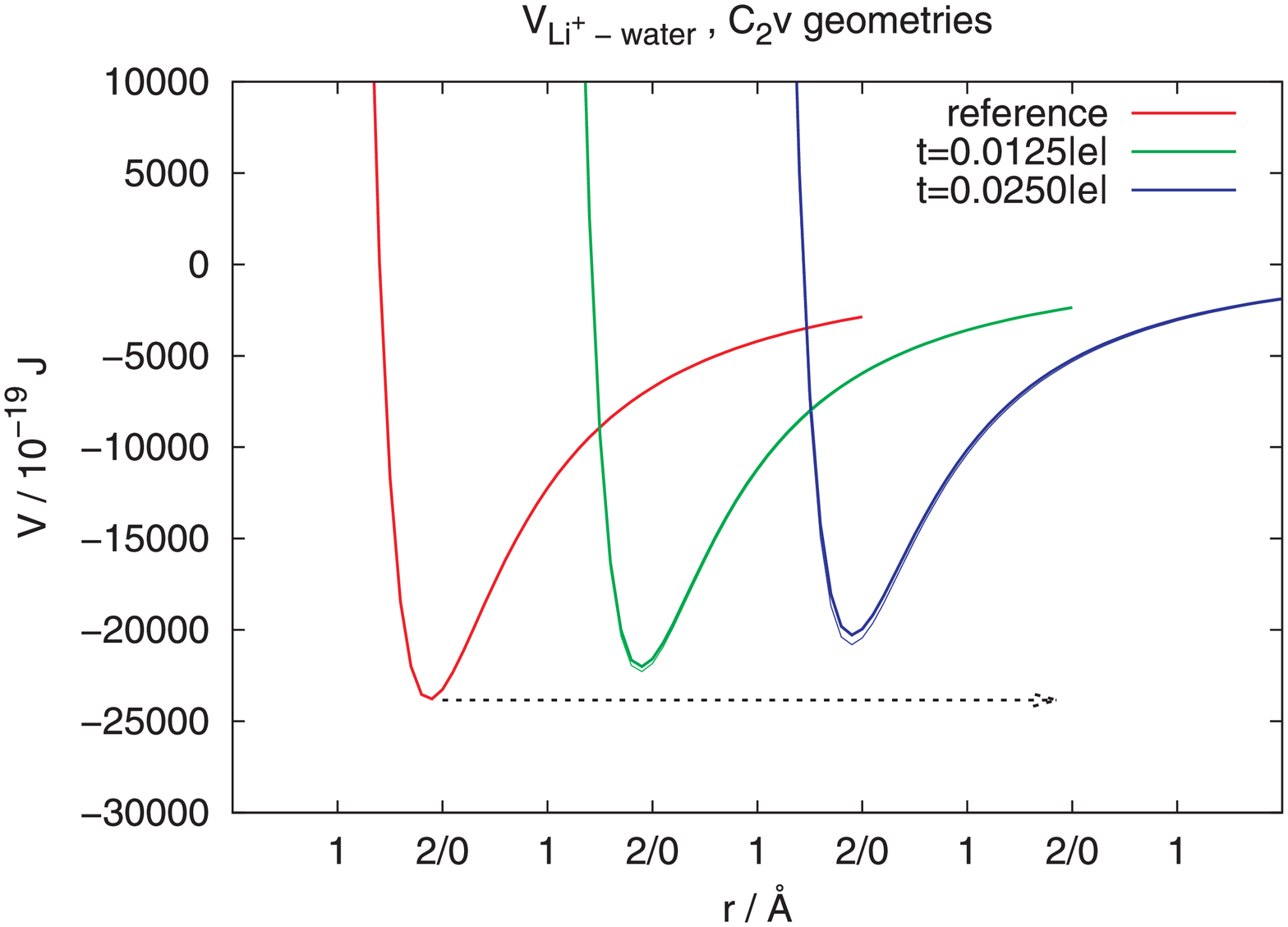}
}
\caption{(Color online) Similar to figure~\ref{pots}: Lithium-water potential wells
for an assumed transfer of
0.0125$|e|$ (green) and 0.0025$|e|$ (blue) from
a Li$^+$-ion to a neighboring (hydration) water molecule.
The fixed geometry of the (flexible) water molecule
is $r_{\rm OH} = 0.9572$~{\AA}, $\angle_{\rm HOH} = 104.5^\circ$, as
in the left curves in
figure~\ref{pots}, and the red curve here is the one for this geometry and
$\alpha=0$ in figure~\ref{pots}.
The lower (thin) green and blue curves
here result form transferring all the charge
from the ion to the oxygen atom of the water molecule,
the upper curves (fat) when one third of the charge is transferred to each
oxygen and hydrogen site of the water model.}
\label{ct}
\end{figure}

Figure~\ref{ftcvv} shows the spectral densities of motion
$d(\nu)$ for the
water molecule oxygens and the Li$^+$-ions in the solutions of lowest and
highest concentrations for the four values of $\alpha$. The spectral density
$d(\nu)$  is
the Fourier-cosine transform of the normalized velocity autocorrelation
function $\bar c_{vv}(t)$:
\[
\bar c_{vv}(t)= c_{vv}(t)/c_{vv}(0) \qquad \text{with} \qquad
c_{vv}(t) = \frac{1}{MN} \sum_{j=1}^M \sum_{i=1}^N
\big[{\mathbf v}_i(\tau_j)\cdot{\mathbf v}_i(\tau_j + t)\big],
\]
\[
d(\nu) \propto \int_0^{t^{\rm corr}}  \bar c_{vv}(t)  \cos{(2 \pi \nu t)}
\ {\rd} t,
\]
where $N$ is the number of equivalent particles (oxygen atoms, Li$^+$-ions),
$M$ is the number of time origins ($t_j$) of the correlation functions,
$\mathbf v$ is the particle velocity and $t^{\rm corr}$ is the length of
the correlation functions (not shown here), here generally 1~ps. Tests with
other correlations lengths (i.e. different numbers $M$ of correlation
origins) yield the same results as the ones shown in the figure.

Figure~\ref{ftcvv} shows that the effect of $\alpha$ on the translational
dynamics is generally minor  here; as expected, the spectral densities are (very
slightly) shifted up  for larger $\alpha$ values.  The effect of the
concentration, on the other hand, is considerable: The spectral densities of
both species show well separated peaks at low concentration which merge more
and more as the concentration increases. The $\nu$=0 term, which is
proportional to the self-diffusion coefficient, also decreases with increasing
concentration. This transition is more or less continuous when the concentration
increases. For this reason, the densities for intermediate concentrations
are not shown here.

\section{Further considerations}

The fact that the polarizabilities of the ion and of the water are not, or only
in an average fashion and not explicitly,  taken into account in the
interaction models has often been criticized, especially if such a model is
to be used in inhomogeneous environments, e.g. at surfaces~\cite{TayDaGa96,DanCha97}.
However, another
effect may also be important and of comparable, if not larger,  magnitude: A
minimal amount of charge transfer $\delta |e|$  (of the order of less than
$\approx $0.03 $|e|$ per water molecule in  the first hydration shell of the
central ion, say) will result in a discharged ion (in our example
Li$^{( 1 - 6 \delta)+}$, on the average) interacting with water molecules charged by the
additional $\delta$, which should be apportioned in some way to the partial
charges of oxygen ($-0.6597|e|$ in the BJH model)  and hydrogen (0.32985$|e|$).
Of course, it is a priori not clear to which site of the model water the charge
is best transferred. As figure~\ref{ct} shows, such a transfer leads in a few
plausible cases (e.g. $\delta |e|$ transferred entirely to the oxygen site of
the water molecule,  or $\delta |e|$ distributed equally among the three water
sites) to markedly shallower ion-water potential wells, comparable to what is
seen in figure~\ref{pots} for different values of $\alpha$. Note that this plot
does not take into account the fact that if several hydration water molecules are
present, the ion will be even more discharged, as described above.

In contrast to the present approach, the water-water interactions between  first shell
water molecules and the other water molecules would also be altered in case of a
transfer of charge. The consequences of this will be explored in future work.
In any case, these simple considerations show how effective
a transfer of charge between an ion and its first shell of solvent molecules
might be in modifying the interactions.

\section{Summary}

We have studied aqueous LiCl solutions in the entire solubility range of this
salt at room temperature and the densities corresponding to ambient pressure. Such
systems are particularly interesting in terms of evaluating the underlying
models because varying the concentration in a wide range means varying   the
contributions originating (in the pair potential approach) from various
kinds of particle pairs (water-water, ion-water,  ion-ion) in an equally wide
range. It was the purpose of this work to show how small modifications
of these interactions affect the results of the simulation. We chose first
to vary the cation-water interaction  since this is where we suspected  in earlier work
a bias leading to an artefact. Besides, we fixed the geometry of the
flexible water model, thus eliminating polarization effects,
which occur mainly in the ionic hydration shell, but also elsewhere.

We have shown that decreasing the Li$^+$-water infraction, in
an admittedly arbitrary fashion, does indeed lead to qualitatively
different results concerning the structure of the solutions, in particular
at intermediate concentrations. Rigidifying the water, which leads to a
reduction both in the ion-water and (less) the water-water
energies, leads to similar, but smaller effects.

This shows how sensitive the results of such
simulations can be to details of the interaction model
and to the balance between the components of such a `global'
model that may have been taken from very different sources
and incorporating different approximations: effective potentials and
ab-initio potentials, rigid vs. flexible models,
polarization, charge transfer, etc.. The theoretical chemists and
solution chemists have their
work cut out.

\section*{Acknowledgements}

Computer time for this project has been provided initially  by P\^ole M3PEC
and, more recently, by the `P\^ole Mod\'elisation' of the department of
chemistry, both at Universit\'e Bordeaux~1.
This work was supported in part
by ``Creating Research Center for Advanced Molecular Biochemistry'',
Strategic Development of Research Infrastructure for
Private Universities, the Ministry of Education, Culture, Sports, Science and
Technology (MEXT), Japan.

\ukrainianpart

\title{Іонні кластери і потенціали іон-вода в МД симуляціях}
\author{Ф.А.  Бопп\refaddr{label1}, К. Ібукі\refaddr{label2}}
\addresses{
\addr{label1} {Університет Бордо, відділ хімії,
 FR--33405 Таланс, Франція}
 \addr{label2}
 {Університет Дошіша, відділ молекулярної хімії та біохімії, Кіото
610--0321, Японія }}

\makeukrtitle

\begin{abstract}
\tolerance=3000%
Добре відомою, хоча і мало описаною, проблемою при виконанні чисельних
молекулярних симуляцій водних іонних розчинів при скінченних
концентраціях є почасти нереалістичні асоціації типу
катіон-катіон. Можна припустити, що причиною цього є дефект іон-іонних
потенціалів взаємодії, про який відомо достатньо мало. Ми показуємо,
що це явище може бути спричинене тим, що при використанні наближення
парної взаємодії потенціали катіон-вода є занадто глибокими порівняно
із іншими. Ми детально досліджуємо цей ефект у нашій роботі та
намагаємось сформувати певні загальні висновки.
\keywords симуляції, молекулярна динаміка, водно-іонні розчини,
іонна асоціація
\end{abstract}


\begin{thebibliography}{99}
%\providecommand{\url}[1]{\texttt{#1}}
%\providecommand{\urlprefix}{URL }
%\providecommand{\eprint}[2][]{\url{#2}}

\bibitem{liem1}
Dang L.X., J. Chem. Phys., 1992, \textbf{96}, 6970; \doi{10.1063/1.462555}.

\bibitem{Lauenstein2000}
Lauenstein A., Hermansson K., Lindgren J., Probst M.M., Bopp P.A., Int. J. Quant. Chem., 2000, \textbf{80}, 892; \doi{10.1002/1097-461X(2000)80:4/5<892::AID-QUA39>3.0.CO;2-Q}.

\bibitem{rode2}
Kritayakornupong C., Plankensteiner K., Rode B.M., J. Comput. Chem., 2004, \textbf{25}, 1576; \doi{10.1002/jcc.20085}.

\bibitem{Pollet2007}
Pollet R., Marx D., J. Chem. Phys., 2007, \textbf{126}, 181102; \doi{10.1063/1.2736369}.

\bibitem{Jardat1999}
Jardat M., Bernard O., Turq P., Kneller G.R., J. Chem. Phys.,
  1999, \textbf{110},  7993; \doi{10.1063/1.478703}.

\bibitem{Ibuki2009}
Ibuki K., Bopp P.A., J. Mol. Liq., 2009, \textbf{147}, 56; \doi{10.1016/j.molliq.2008.08.005}.

\bibitem{Turton2011}
Turton D.A., Corsaro C., Candelaresi M., Brownlie A., Seddon K.R., Mallamace
  F., Wynne K., Faraday Discuss., 2011, \textbf{150}, 493; \doi{10.1039/c0fd00005a}.

\bibitem{Terekhova2010}
Terekhova I.V., Romanova A.O., Kumeev R.S., Fedorov M.V., J. Phys. Chem. B, 2010, \textbf{114}, 12607;\\  \doi{10.1021/jp1063512}.

\bibitem{Groenenboom2000}
Groenenboom G., Mas E., Bukowski R., Szalewicz K., Wormer P., van der Avoird~A., Phys. Rev. Lett., 2000, \textbf{84}, 4072; \doi{10.1103/PhysRevLett.84.4072}.

\bibitem{Mas2000}
Mas E.M., Bukowski R., Szalewicz K., Groenenboom G.C., Wormer P.E.S., van~der~Avoird~A., J. Chem. Phys., 2000, \textbf{113},  6687; \doi{10.1063/1.1311289}.

\bibitem{Guillot2001}
Guillot B., Guissani Y., J. Chem. Phys., 2001, \textbf{114},
   6720; \doi{10.1063/1.1356002}.

\bibitem{Guill02}
Guillot B., J. Mol. Liq., 2002, \textbf{101},  219; \doi{10.1016/S0167-7322(02)00094-6}.

\bibitem{Pusztai2008}
Pusztai L., Pizio O., Sokolowski S., J. Chem. Phys., 2008,
  \textbf{129},  184103; \doi{10.1063/1.2976578}.

\bibitem{Harsanyi2011}
Hars\'{a}nyi I., Bopp P.A., Vrhov\v{s}ek A., Pusztai L., J. Mol. Liq., 2011, \textbf{158}, 61; \doi{10.1016/j.molliq.2010.10.010}.

\bibitem{firstnacl}
Bopp P.A., Dietz W., Heinzinger K., Z. Naturforsch., 1979,
  \textbf{34a}, 1424.

\bibitem{bjh-intra}
Bopp P.A., Jancs\'{o} G., Heinzinger K., Chem. Phys. Lett., 1983,
  \textbf{98}, 129; \doi{10.1016/0009-2614(83)87112-7}.

\bibitem{Kast1996}
Kast S.M., Brickmann J., J. Chem. Phys., 1996, \textbf{104},
   3732; \doi{10.1063/1.471028}.

\bibitem{TayDaGa96}
Taylor R.S., Dang L.X., Garrett B.C., J. Phys. Chem., 1996,
  \textbf{100}, 11720; \doi{10.1021/jp960615b}.

\bibitem{DanCha97}
Dang L.X., Chang T.M., J. Chem. Phys., 1997, \textbf{106},
   8149; \doi{10.1063/1.473820}.

\end{thebibliography}
\end{document}